\documentclass[11pt,draft]{article}
\pdfoutput=1
\usepackage[final]{graphicx}            %
\DeclareGraphicsExtensions{.pdf, .jpg}  %

\usepackage{amsmath}                    %
\usepackage{amssymb}                    %
\usepackage{mathtools}                    %
\usepackage{amsthm}
\usepackage{mathrsfs}                   %
\usepackage{stmaryrd}                   %
\usepackage[symbol,perpage]{footmisc}
\usepackage[sort]{natbib}               %
\usepackage[np,autolanguage]{numprint}
\usepackage{color}
\usepackage{afterpage}
\usepackage{setspace}
\usepackage{enumerate}
\theoremstyle{plain}                    %
\theoremstyle{definition}               %
\theoremstyle{remark}                   %
\newtheorem{remark}{Remark}             %
\renewcommand{\d}[1]{\ensuremath{\mathrm{d}#1}}
\newcommand{\bv}[1]{\boldsymbol{#1}}    % 
\newcommand{\uv}[1]{\hat{\boldsymbol{u}}_{#1}}       %
\newcommand{\tensor}[1]{\mathsf{#1}}    %
\newcommand{\trans}[1]{#1^{\mathrm{T}}} %
\newcommand{\invtrans}[1]{\ensuremath{#1^{-\mathrm{T}}}} %

\newcommand{\FigPath}[1]{#1}
\DeclareMathOperator{\trace}{tr}        % 
\usepackage[paper = letterpaper,
            lmargin = 1in, rmargin = 1in,
            tmargin = 1in, bmargin = 1in]{geometry}

\title{%
The Invalidity of the Laplace Law for Biological Vessels and of Estimating Elastic Modulus from Total Stress vs.\ Strain: a New Practical Method%
\footnote{Submitted for publication in \emph{Mathematical Medicine \& Biology} (a journal of the IMA) on 4 February 2013, manuscript ID: MMB-13-005.}
}

\author{%
Francesco Costanzo\\
Center for Neural Engineering\\
The Pennsylvania State University\\
W-315 Millennium Science Complex\\
University Park,
PA 16802,
USA
\and
James G. Brasseur\\
Department of Mechanical and Nuclear Engineering\\
The Pennsylvania State University\\
205 Reber Building\\
University Park,
PA 16802,
USA%
}
\begin{document}
%%%%%%%%%%%%%%%%%%%%%%%%%%%%%%
% *** BEGIN  FRONTMATTER *** %
%%%%%%%%%%%%%%%%%%%%%%%%%%%%%%

\maketitle

%: Abstract
\begin{abstract}
The quantification of the stiffness of tubular biological structures is often obtained, both \emph{in vivo} and \emph{in vitro}, as the slope of total transmural hoop stress plotted against hoop strain.  Total hoop stress is typically estimated using the ``Laplace law.''  We show that this procedure is fundamentally flawed for two reasons: Firstly, the Laplace law predicts total stress incorrectly for biological vessels.  Furthermore, because muscle and other biological tissue are closely volume-preserving, quantifications of elastic modulus require the removal of the contribution to total stress from incompressibility.  We show that this hydrostatic contribution to total stress has a strong material-dependent nonlinear response to deformation that is difficult to predict or measure. To address this difficulty, we propose a new practical method to estimate a mechanically viable modulus of elasticity that can be applied both \emph{in vivo} and \emph{in vitro} using the same measurements as current methods, with care taken to record the reference state. To be insensitive to incompressibility, our method is based on shear stress rather than hoop stress, and provides a true measure of the elastic response without application of the Laplace law.  We demonstrate the accuracy of our method using a mathematical model of tube inflation with multiple constitutive models. We also re-analyze an \emph{in vivo} study from the gastro-intestinal literature that applied the standard approach and concluded that a drug-induced change in elastic modulus depended on the protocol used to distend the esophageal lumen.  Our new method removes this protocol-dependent inconsistency in the previous result.
\end{abstract}

%%%%%%%%%%%%%%%%%%%%%%%%%%%%%%%%%%%%%%%%%%%%%%%%%%
% This is an AMS-LaTeX command to allow breaking %
% of displayed equations across pages. Note the  %
% closing the "}" just before the bibliography.  %
%%%%%%%%%%%%%%%%%%%%%%%%%%%%%%%%%%%%%%%%%%%%%%%%%%
\allowdisplaybreaks{                             %
%%%%%%%%%%%%%%%%%%%%%%%%%%%%%%%%%%%%%%%%%%%%%%%%%%%
%%: Introduction
\section{Introduction}
\label{section: introduction}
Although originally developed to estimate mathematically the surface tension at the air-liquid interface in bubbles as a function of geometry and bubble pressure, the ``Laplace law'' is used widely in the medical community to estimate the average hoop stress in the wall of biological vessels such as blood vessels and the muscle-driven conduits of the gastrointestinal (GI) tract \citep{Basford_2002_The-Law-of-Laplace_0}.  In the GI tract, for example, the Laplace law is commonly used to estimate average hoop stress across the esophageal wall\footnote{The esophageal wall is composed of circular and longitudinal muscle layers, and inner mucosal layers of mostly loosely connected tissue.} as a function of distension and manometrically measured pressure within the esophageal lumen \citep{BiancaniZabinski_1975_Pressure_0,GhoshKahrilas_2005_The-Mechanical_0,FrokjaerAndersen_2006_Ultrasound-determined_0,GregersenGilja_2002_Mechanical_0,PedersenDrewes_2005_New-analysis_0,TakedaNabae_2004_Oesophageal_0,TakedaKassab_2002_A-novel_0,TakedaKassab_2003_Effect_0,YangZhao_2004_Biomechanical_0,YangFung_2006_Directional_0,YangFung_2006_3D-mechanical_0,YangFung_2006_Viscoelasticity_0,YangFung_2007_Instability_0}.  The Laplace law is also commonly applied in the cardiovascular system (see, e.g., \citealp{Kassab2006Biomechanics-of0}). 

In addition, to estimate material stiffness, the wall-average hoop stress of the vessel wall is commonly plotted against a stretch measure and the slope of the curve interpreted as a modulus of elasticity or wall stiffness.\footnote{Here the word ``stiffness'' is used, as in the medical literature, to mean ``material resistance to deformation.''} For example, \cite{TakedaNabae_2004_Oesophageal_0,TakedaKassab_2003_Effect_0} determine a modulus of elasticity by plotting the circumferential second Piola-Kirchhoff stress against the circumferential component of the Green strain,\footnote{This strain is also called the Green-Saint Venant strain\citep{GurtinFried_2010_The-Mechanics_0}, the Green-Lagrange strain  \citep{Holzapfel-CMBook-2000-1}, or the Lagrangian strain \citep{Bowen-CMBook-1989-1,Taber_2004_Nonlinear_0}.} and \cite{GregersenGilja_2002_Mechanical_0} plot the Cauchy circumferential stress against a circumferential stretch ratio, where in both cases stress is estimated using the Laplace law.  A major advantage of this approach is that, with the Laplace law, hoop stress and deformation can be estimated, \emph{in vivo} as well as \emph{in vitro}, from measurements of intraluminal pressure and luminal cross sectional area with concurrent manometry and endoluminal ultrasound \citep{GregersenGilja_2002_Mechanical_0,TakedaNabae_2004_Oesophageal_0,TakedaKassab_2002_A-novel_0,TakedaKassab_2003_Effect_0,YangZhao_2004_Biomechanical_0}.  With this approach, luminal wall stiffness is estimated and compared across different clinical groups such as patient/healthy, old/young and pre/post therapy, or over time.

We show that this approach to measuring a modulus of elasticity is flawed for two reasons. Firstly, we show that the Laplace law is invalid and in serious error as an estimate of total hoop stress when applied to biological vessels with physiological deformations. Secondly, we demonstrate that the slope of average total hoop stress plotted against an appropriate strain measure does not represent a modulus of elasticity, both in principle and in practice, because, to measure stiffness, the hydrostatic component of the stress due to incompressibility must be removed.  Clearly, there remains a great need for a practical approach to estimate average material stiffness of biological vessels \emph{in vivo} and \emph{in vitro}. We propose an alternative strategy with which a true modulus of elasticity can be estimated from data obtained via current \emph{in vivo} and \emph{in vitro} experimental methods and we demonstrate the validity of our new method using parameters for the esophagus.

\section{The Laplace law as an estimate of average hoop stress in biological vessels}
\label{sec: LL estimate of total Hoop stress}
Wall stiffness of biological vessels is commonly estimated by interpreting the slope of a total hoop stress against strain as an elastic modulus.  Both \emph{in vivo} and \emph{in vitro}, the raw data used to generate the curve consist of measurements of the transmural pressure difference and of the inner and outer radii of the vessel.  The Laplace law is then used to translate these measurements into the needed wall-average total hoop stress.  In this study we address the following questions:
\begin{enumerate}
\item
Since the Laplace law is an approximation valid in the limit of thin-walled vessels, how well does the Laplace law approximate the total average hoop stress in biological vessels?

\item
Since muscle and other biological tissue are nearly incompressible and the total hoop stress has both an elastic component and a component due to volume conservation (due to incompressibility), does the total average hoop stress (plotted against strain) approximate well the elastic response of the vessel (relative to the purely elastic component)?
\end{enumerate}
We address these questions by first reviewing the derivation of the Laplace law; then comparing the average hoop stress estimates obtained via the Laplace law with exact estimates based on two rigorous solutions.

\subsection{The Laplace law}
\label{subsec: The Laplace Law}
The Laplace law represents an equilibrium force balance across the wall of a tubular structure with an infinitesimally thin wall. Consider in Fig.~\ref{fig: coordinate systems}
%%%
\begin{figure}[htb]
    \centering
    \includegraphics{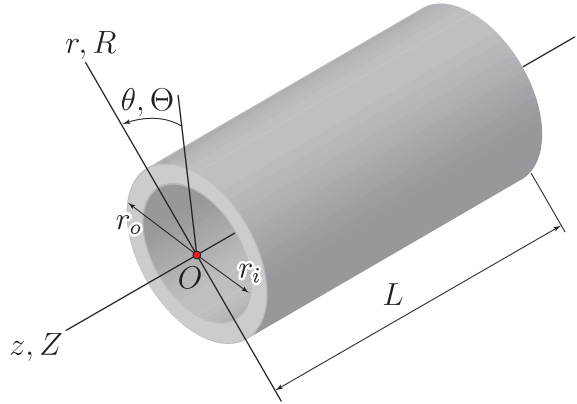}
    \caption{A circular cylinder and the coordinate systems used to describe the reference and deformed configurations.}
    \label{fig: coordinate systems}
\end{figure}
%%%
a circular cylinder that is distended in response to a transmural pressure difference $\Delta p = p_{i} - p_{o}$ across the wall, where $p_{i}$ and $p_{o}$ are the inner and outer pressures, respectively.  The cylinder's inner and outer radii are $r_{i}$ and $r_{o}$, respectively, so that the wall's thickness is $t = r_{o} - r_{i}$.  If the geometry of the cylinder is axisymmetric and the material is isotropic, the Cauchy hoop stress $\sigma_{\theta\theta}$ depends only on the radial coordinate $r$ and not on the axial coordinate $z$ or azimuthal angle $\theta$.  Denoting by $\langle \bullet \rangle$ the average of a quantity across the cylinder wall, the average hoop stress is
\begin{equation}
\label{Eq: average operator def}
\langle \sigma_{\theta\theta} \rangle = \frac{1}{r_{o} - r_{i}} \int_{r_{i}}^{r_{o}} \sigma_{\theta\theta} \, \d{r}.
\end{equation}
Cutting the cylinder longitudinally as illustrated in Fig.~\ref{fig: half cylinder FBD},
%%%
\begin{figure}[htb]
    \centering
    \includegraphics{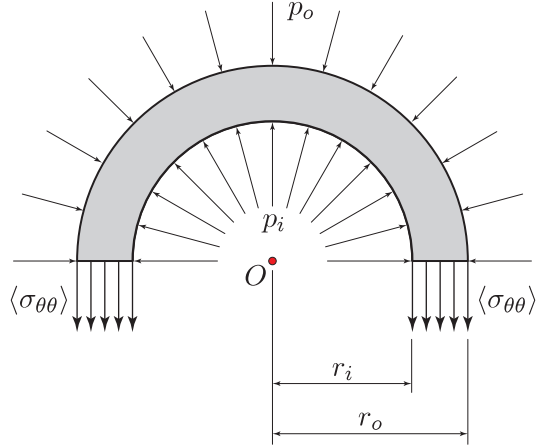}
    \caption{Free-body diagram of half of a cross section of the pressurized cylinder.  The average force distribution acting on the cut is shown.}
    \label{fig: half cylinder FBD}
\end{figure}
%%%
and balancing the resultant force per unit length due to $p_{i}$ over half the cylinder in the direction perpendicular to the cut (with magnitude $2 p_{i} r_{i}$) with the corresponding resultant force due to $p_{o}$ and $\langle\sigma_{\theta\theta}\rangle$ (with magnitude $2 p_{o} r_{o} + 2 \langle\sigma_{\theta\theta}\rangle t$) one obtains:
\begin{equation}
\label{Eq: Laplace law Full}
\langle \sigma_{\theta\theta} \rangle = \frac{\Delta p}{t/r_{i}} - p_{o}.
\end{equation}
The Laplace law is the limit of Eq.~\eqref{Eq: Laplace law Full} for $t/r_{i} \to 0$.  For bounded values of $p_{o}$ and $p_{i}$, the term $\Delta p/(t/r_{i})$ becomes dominant relative to $p_{o}$ and the equilibrium force balance for a ``thin walled'' cylinder, the ``Laplace law,'' becomes
\begin{equation}
\label{Eq: Laplace law intro}
\langle \sigma_{\theta\theta} \rangle_{\text{LL}} = \frac{\Delta p}{t/r_{i}}.
\end{equation}
The thin wall approximation, typically expressed as $t/r_{i} \ll 1$, could be naively interpreted in practice by taking $t/r_{i} < 0.1$. However, the passage from Eq.~\eqref{Eq: Laplace law Full} to Eq.~\eqref{Eq: Laplace law intro} requires that $p_{o} \ll \Delta p/(t/r_{i})$, leading to the following requirement for a proper use of the Laplace law:
\begin{equation}
\label{Eq: Justification for use of LL}
\frac{t}{r_{i}} \ll \frac{\Delta p}{p_{o}}.
\end{equation}
This condition is rarely met in biological vessels. For example, radial distention of the esophagus during the transport of food boluses between the pharynx and stomach are driven by pressure differences up to a maximum of  $\approx 25\text{\,--\,}\np[mmHg]{30}$ \citep{GhoshKahrilas_2005_The-Mechanical_0}. Therefore, since $p_{o}$ is close to atmospheric pressure, for the Laplace law to be applicable in the esophagus, the relative thickness of the esophageal wall must be extremely small: $t/r_{i} \ll 0.04$.  This condition is almost never met in the human body. For example, extensive measurements of esophageal cross sections using endoluminal ultrasound before, during, and after esophageal distension with bolus transport \citep{NicosiaBrasseur_2001_Local_0,Schiffner_2004_Opening_0,UlerichDai_2003_Detailed_0} indicate that $t/r_{i}$ typically ranges between a maximum value $(t/r_{i})_\text{max} \approx 0.7$, in the pre-swallow resting state, to a minimum value $(t/r_{i})_\text{min} \approx 0.1$ in the most distended state. Thus the condition $t/r_{i} \ll 0.04$  is never met, not even approximately. 

Analysis in blood vessels leads to a similar conclusion.  The transmural pressure difference $\Delta p$, for example, is approximately $\np[mmHg]{100}$ or less, depending on the vessel and degree of hypertension (\citealp{Kim2002Three-Dimension0,Pries2005Remodeling-of-B0,Heagerty1993Small-Artery-St0}), so that $\Delta p/p_{o} \approx 0.13$ or less. Thus, $t/r_{i}$ must be much smaller than $0.13$ for the Laplace law to be a reasonable approximation. \cite{Kim2002Three-Dimension0}, using MRI, report normal $t/r_{i}$ ratios in the coronary arteries around $0.5$ in humans and much higher with hypertension. \cite{Pries2005Remodeling-of-B0} report values of $t/r_{i}$ from the literature for arterioles in rats $\approx 0.21\text{\,--\,}0.26$, while \cite{Heagerty1993Small-Artery-St0} report \emph{in vitro} values of normal small arteries in humans $\approx 0.10\text{\,--\,}0.13$, and for hypertensive small arteries $\approx 0.16\text{\,--\,}0.20$. None of these values come close to the requirement that $t/r_{i}$ be  $\ll 0.13$ for the Laplace law to be an accurate approximation of total hoop stress.

With the above in mind, we wish to provide a quantitative assessment of the accuracy of the Laplace law  in incompressible materials.  From this assessment, we can then present an objective critique of the practice of measuring a modulus of elasticity from a plot of such average stress against strain.  This goal can be readily achieved by determining an exact solution of the inflation and extension problem for an elastic tubular vessel. Here, we considered two that are particularly simple: the first for an isotropic neo-Hookean material, and the second for a simplified Mooney-Rivlin material.  In both cases, the constitutive model is chosen so to have a single constitutive parameter that could be estimated using the information measured \emph{in vivo}.  Whether or not the chosen models are appropriate for the modeling of biological tissue is a secondary concern here.  The analysis presented here is meant to show in a quantitative way that current practices for the determination of tissue stiffness are inadequate and that the development of more rigorous methods for the task is justified.

\subsection{Inflation and Extension of Tubular Vessels: Exact Solutions for Two Incompressible Isotropic Materials}
\label{subsec: Inflation and Extension of a Circular Cylinder}
To evaluate the accuracy of the Laplace law for biological vessels, we derived \emph{exact} expressions for the stress field in an incompressible cylindrical vessel subject to prescribed axial stretch $\lambda_{z}$ and transmural pressure difference $\Delta p$.  Two different models were chosen to show that the same conclusions are reached in both cases.  These models were also chosen so as to be able to derive solutions in closed form and no claims are made about their ability to represent the behavior of a particular biological tissue.

Solutions to the inflation and extension of a circular cylinder has been discussed by various authors (see, for example, \citealp{Ogden-NEDBook-1984-1,Humphrey_2010_Cardiovascular_0}; see also the analysis by \citealp{De-Pascalis_2010_The-Semi-Inverse_0}).  However, the present authors are not aware of solutions that show in detail the full stress field in the form needed for the analysis at hand.  For this reason and for the sake of completeness, the relevant details of our solution are reported in the Appendix.

%\subsection{Kinematics}
%\label{subsec: kinematics}
\paragraph{Kinematics.}
Consider the inflation and extension of a circular cylinder with uniform wall thickness (Fig.~\ref{fig: coordinate systems}).  Points in the reference configuration are identified by the cylindrical coordinate coordinates $(R, \Theta, Z)$ (the origin is on the longitudinal axis of the cylinder) ranging as follows:
\begin{equation}
\label{Eq: RC Domain}
R_{i} < R < R_{o},
\quad
0 < \Theta < 2 \pi,
\quad \text{and} \quad
0 < Z < L,
\end{equation}
where $R_{i}$, $R_{o}$, and $L$ are the internal radius, external radius, and length of the cylinder, respectively.  As discussed later in detail, we assume that the reference configuration is \emph{unloaded} in the sense that the pressure difference across the cylinder's wall is equal to zero.  Points in the deformed configuration are identified by the coordinates $(r, \theta, z)$.  By inflation and extension we mean a deformation such that
\begin{equation}
\label{Eq: Motion specification}
r = r(R),
\quad
\theta = \Theta,
\quad
z = \lambda_{z} Z,
\end{equation}
where $r(R)$ is at least twice differentiable.  The quantity $\lambda_{z} > 0$, uniform across the wall, is the prescribed longitudinal stretch of the cylinder.  We denote by $r_{i}$ and $r_{o}$ the values of $r$ for $R = R_{i}$ and $R = R_{o}$, respectively.  We use two sets of orthonormal base vectors, namely $\bigl(\uv{R}, \uv{\Theta}, \uv{Z} \bigr)$ and $\uv{r}, \uv{\theta}, \uv{z} \bigr)$,
%%
%\begin{equation}
%\label{Eq: UV Ref}
%\bigl(\uv{R}, \uv{\Theta}, \uv{Z} \bigr)
%\quad \text{and} \quad
%\bigl(\uv{r}, \uv{\theta}, \uv{z} \bigr),
%\end{equation}
%%
tangent to the corresponding coordinate lines of the $(R, \Theta, Z)$ and $(r, \theta, z)$ coordinate systems, respectively.  When the material is incompressible the volume is preserved so that we must have
\begin{equation}
\label{Eq: continuity}
r' \frac{r}{R} \lambda_{z} = 1
\quad \Rightarrow \quad
\lambda_{z} \bigl(r^{2} - r_{i}^{2}\bigr) = R^{2} - R_{i}^{2},
\end{equation}
where $\bullet' = \d{\bullet}/\d{R}$, and where $r = r_{i}$ for $R = R_{i}$.  From the second of Eqs.~\eqref{Eq: continuity} we see that, when $\lambda_{z} = 1$ (i.e., no longitudinal shortening or extension), the cross-sectional area of of the cylinder is also preserved during deformation.

For future reference, the deformation gradient $\tensor{F}$, $\tensor{F}^{-1}$, the left Cauchy-Green strain tensors $\tensor{B}$, and its inverse $\tensor{B}^{-1}$ are:
\begin{align}
\label{Eq: F}
%\tensor{F} &= r' \, \uv{r} \otimes \uv{R} + \frac{r}{R} \, \uv{\theta} \otimes \uv{\Theta} + \lambda_{z} \, \uv{z} \otimes \uv{Z},
\tensor{F} &= \frac{1}{\lambda_{z}} \frac{R}{r} \, \uv{r} \otimes \uv{R} + \frac{r}{R} \, \uv{\theta} \otimes \uv{\Theta} + \lambda_{z} \, \uv{z} \otimes \uv{Z},
\\
\label{Eq: FInv}
%\tensor{F} &= \frac{1}{r'} \, \uv{R} \otimes \uv{r}  + \frac{R}{r} \, \uv{\Theta} \otimes \uv{\theta}   + \frac{1}{\lambda_{z}} \uv{Z} \otimes  \, \uv{z},
\tensor{F}^{-1} &= \lambda_{z} \frac{r}{R} \, \uv{R} \otimes \uv{r}  + \frac{R}{r} \, \uv{\Theta} \otimes \uv{\theta}   + \frac{1}{\lambda_{z}} \uv{Z} \otimes  \, \uv{z},
\\
\label{Eq: B}
\tensor{B} &= \frac{1}{\lambda_{z}^{2}} \frac{R^{2}}{r^{2}}\, \uv{r} \otimes \uv{r} + \frac{r^{2}}{R^{2}} \, \uv{\theta} \otimes \uv{\theta} + \lambda_{z}^{2} \, \uv{z} \otimes \uv{z},
\\
\label{Eq: B inverse}
\tensor{B}^{-1} &= \lambda_{z}^{2} \frac{r^{2}}{R^{2}}\, \uv{r} \otimes \uv{r} + \frac{R^{2}}{r^{2}} \, \uv{\theta} \otimes \uv{\theta} + \frac{1}{\lambda_{z}^{2}} \, \uv{z} \otimes \uv{z}.
\end{align}
%%
%The corresponding principal stretches in the radial, hoop, and longitudinal directions, respectively, are
%%%
%\begin{equation}
%\label{Eq: Principal stretches}
%\lambda_{1} = \lambda_{r} = r' = \frac{1}{\lambda_{z}}\frac{R}{r},
%\quad
%\lambda_{2} = \lambda_{\theta} = \frac{r}{R},
%\quad
%\lambda_{3} = \lambda_{z}.
%\end{equation}
%%%

%\subsection{Stress solutions for two incompressible isotropic materials}
%\label{subsec: stress solution}
\paragraph{Stress field.}
To determine the stress field in the deformed state it is necessary to define the material through constitutive equations.  As mentioned earlier, our choice of constitutive equations is motivated by our desire to quantify the accuracy of the Laplace law as an estimate for the total average hope stress.  We choose two constitutive relations from which \emph{full and exact} stress solutions can be obtained: a neo-Hookean model and a simplified Mooney-Rivlin model.  By modeling the esophagus, for example, as a homogeneous, isotropic, and incompressible hyperelastic material, we neglect rate dependence \citep{YangFung_2006_Viscoelasticity_0}, the layered structure of the esophagus \citep{NicosiaBrasseur_2002_A-Mathematical_0,GhoshKahrilas_2008_Liquid_0,YangFung_2006_3D-mechanical_0,YangFung_2006_Directional_0,YangFung_2007_Instability_0}, and the potential existence of heterogeneity and non-hydrostatic residual stress fields.

Since average intrathoracic pressure is approximately atmospheric, and because the esophageal reference state has been taken as the unloaded, the inner and outer pressures $p_{i}$ and $p_{o}$ are approximately equal, in the reference state we set $p_{i}$ and $p_{o}$ equal to atmospheric pressure.  In addition, in our analysis we consider equilibrium states under negligible body forces.
%
%We denote by $\bv{\sigma}$ and $\tensor{S}$ the Cauchy and the second Piola-Kirchhoff stress tensors, respectively.  The tensors $\tensor{S}$ is related to $\bv{\sigma}$ as follows:
%%%
%\begin{equation}
%\label{Eq: T-S relation}
%%\tensor{P} = (\det \tensor{F}) \bv{\sigma} \invtrans{\tensor{F}}
%%\quad \text{and} \quad 
%\tensor{S} = (\det \tensor{F}) \tensor{F}^{-1} \bv{\sigma} \invtrans{\tensor{F}}.
%\end{equation}
%%
Because the material is incompressible, if all of the material constitutive parameters are known and the (volume-preserving) deformation prescribed, the stress solution is completely known to within a Lagrange multiplier $q$, which is found by enforcing static equilibrium and boundary conditions.  In addition, limiting our attention to incompressible isotropic materials with undistorted reference configurations\footnote{The main implication of this assumption is that if a residual stress field exists, this field must be hydrostatic.} \citep{Bowen-CMBook-1989-1}, the only nontrivial equilibrium equation of the problem, written in terms of the Cauchy stress, has the following form: 
\begin{equation}
\label{Eq: NHs div T = 0}
\frac{\d{\sigma_{rr}}}{\d{r}} + \frac{\sigma_{rr} - \sigma_{\theta\theta}}{r} = 0.
\end{equation}
%%

%\subsubsection{Homogeneous incompressible neo-Hookean isotropic material}
\paragraph{Neo-Hookean material.}
The Cauchy stress constitutive response function for a homogeneous isotropic neo-Hookean material with undistorted reference configuration is
\begin{equation}
\label{Eq: Neohookean Material}
\bv{\sigma} = -\bar{q} \tensor{I} + \mu (\tensor{B} - \tensor{I}),
\end{equation}
where $\bar{q}$ is the Lagrange multiplier required for the satisfaction of incompressibility.  In our problem, $\tensor{B}$ is given by Eq.~\eqref{Eq: B}.  Since $\mu$ is constant, we absorb it into $\bar{q}$ and rewrite Eq.~\eqref{Eq: Neohookean Material} with the substitution $q = \bar{q} + \mu$:
\begin{equation}
\label{Eq: Effective stress strain}
\bv{\sigma} = -q \tensor{I} + \mu \tensor{B}.
%\quad \text{and} \quad
%\tensor{P} = - q \invtrans{\tensor{F}} + \mu \tensor{F},
\end{equation}
Enforcing static equilibrium and boundary conditions, we find the following expression for $q$ (see Eq.~\eqref{Eq: q final form} in the Appendix):
\begin{equation}
\label{Eq: qNH solution}
%q = p_{o} + \frac{\Delta p}{\gamma(r_{i},r_{o},R_{i},R_{o},\lambda_{z})} \biggl[
%\frac{1}{2 \lambda_{z}^{2}} 
%\biggl(\frac{R_{o}^{2}}{r_{o}^{2}} + \frac{R^{2}}{r^{2}} \biggr)
%+ \frac{1}{\lambda_{z}} \ln\biggl(\frac{r R_{o}}{r_{o} R}\biggr)
%\biggr],
(q)_{\text{NH}} = p_{o} + \mu \biggl[
\frac{1}{2 \lambda_{z}^{2}} 
\biggl(\frac{R_{o}^{2}}{r_{o}^{2}} + \frac{R^{2}}{r^{2}} \biggr)
+ \frac{1}{\lambda_{z}} \ln\biggl(\frac{r R_{o}}{r_{o} R}\biggr)
\biggr],
\end{equation}
where the subscript `$\text{NH}$' stands for `neo-Hookean.'  The quantity $\mu$ can be expressed in terms of quantities that are measured and/or controlled in an experiment (see Eq.~\eqref{Eq: mu solution} in the Appendix):
\begin{gather}
\label{Eq: mu bar solution}
\mu = \frac{\Delta p}{\gamma(r_{i},r_{o},R_{i},R_{o},\lambda_{z})}
\\
\shortintertext{with}
\label{Eq: gamma def}
\gamma(r_{i},r_{o},R_{i},R_{o},\lambda_{z}) = \frac{1}{2 \lambda_{z}^{2}} \biggl( \frac{R_{o}^{2}}{r_{o}^{2}} - \frac{R_{i}^{2}}{r_{i}^{2}} \biggr) + \frac{1}{\lambda_{z}} \ln\biggl( \frac{r_{i} R_{o}}{r_{o} R_{i}} \biggr).
\end{gather}
%%
%Combining Eqs.~\eqref{Eq: sigmabar function of Delta p}--\eqref{Eq: gamma def}, we obtain an expression for $q$ in terms of those quantities that are measured and/or controlled in an experiment:
%%%
%\begin{equation}
%\label{Eq: sigmabar function of experimental data}
%(q)_{\text{NH}} = p_{o} + \Delta p
%\frac{
%\frac{1}{2 \lambda_{z}^{2}} 
%\left(\frac{R_{o}^{2}}{r_{o}^{2}} + \frac{R^{2}}{r^{2}} \right)
%+ \frac{1}{\lambda_{z}} \ln\left(\frac{r R_{o}}{r_{o} R}\right)
%}
%{
%\frac{1}{2 \lambda_{z}^{2}} \left( \frac{R_{o}^{2}}{r_{o}^{2}} - \frac{R_{i}^{2}}{r_{i}^{2}} \right) + \frac{1}{\lambda_{z}} \ln\left( \frac{r_{i} R_{o}}{r_{o} R_{i}} \right)
%}.
%\end{equation}
%%
Combining Eqs.~\eqref{Eq: B} and~\eqref{Eq: qNH solution} into Eq.~\eqref{Eq: Effective stress strain}, we have that the hoop stress $\sigma_{\theta\theta}$ for the neo-Hookean material is given by (see Eq.~\eqref{Eq: sigma final app} in the Appendix)
\begin{equation}
\label{Eq: sigmathetathetaNH solution}
(\sigma_{\theta\theta})_{\text{NH}}
=
-p_{o}
- \mu \left[\frac{1}{2 \lambda_{z}^{2}} 
\left(\frac{R_{o}^{2}}{r_{o}^{2}} + \frac{R^{2}}{r^{2}} \right)
+ \frac{1}{\lambda_{z}} \ln\left(\frac{r R_{o}}{r_{o} R}\right)\right]
+ \mu \frac{r^{2}}{R^{2}}.
\end{equation}
%%
%Next, using the above result along with Eqs.~\eqref{Eq: FInv} and~\eqref{Eq: T-S relation}, keeping in mind that, for an incompressible material, $\det\tensor{F} = 1$, we have that the hoop component of the second Piola-Kirchhoff stress tensor is
%%%
%\begin{equation}
%\label{Eq: SthetathetaNH solution}
%(S_{\Theta\Theta})_{\text{NH}}
%=
%-p_{o} \frac{R^{2}}{r^{2}}
%- \mu \left[\frac{1}{2 \lambda_{z}^{2}} 
%\left(\frac{R_{o}^{2}}{r_{o}^{2}} + \frac{R^{2}}{r^{2}} \right)
%+ \frac{1}{\lambda_{z}} \ln\left(\frac{r R_{o}}{r_{o} R}\right)\right]
%\frac{R^{2}}{r^{2}}
%+ \mu
%\end{equation}
%%%
In practice, the empirical measurements of wall stiffness are made using wall-average quantities.  This is precisely the case when the Laplace law is used since this law provides an estimate of average of the hoop stress across a vessel's wall.  For this reason, we now use the pointwise results in Eqs.~\eqref{Eq: qNH solution} and~\eqref{Eq: sigmathetathetaNH solution} to obtain their average across the wall of the cylinder.  Specifically, letting $\rho = \sqrt{r_{i}^{2} - R_{i}^{2}/\lambda_{z}} = \sqrt{r_{o}^{2} - R_{o}^{2}/\lambda_{z}}$, and using Eqs.~\eqref{Aeq: r2R2 and R2r2}--\eqref{Aeq: R4r4 log1 and rRlog avg3} in the Appendix and simplifying, we have
\begin{align}
\label{Eq: AvgqNH}
\langle q \rangle_{\text{NH}} &= p_{o}
+
\frac{\mu}{2 \lambda_{z}^{2}} \frac{R_{o}^{2}}{r_{o}^{2}}
+ \frac{\mu}{2 \lambda_{z}} 
\biggl(1 - \frac{\rho^{2}}{r_{o}r_{i}} \biggr)
\notag
\\
&\quad+
\frac{\mu}{2 \lambda_{z}(r_{o} - r_{i})} \biggl[ 
\rho \ln
\frac{(r_{o} - \rho)(r_{i} + \rho)}{(r_{o} + \rho)(r_{i} - \rho)}
- 2 r_{i} \ln\biggl(\frac{r_{i} R_{o}}{r_{o}R_{i}}\biggr)
\biggr],
\\
\label{Eq: AvgsigmathetathetaNH}
\langle\sigma_{\theta\theta}\rangle_{\text{NH}} &= 
-p_{o}
+
\frac{\mu r_{i}}{2 \lambda_{z}^{2} (r_{o} - r_{i})}
\Biggl[
\frac{R_{o}^{2}}{r_{o}^{2}} - \frac{R_{i}^{2}}{r_{i}^{2}}
+
2 \lambda_{z}
\ln\biggl(\frac{r_{i} R_{o}}{r_{o}R_{i}}\biggr)
\Biggr],
%\\
%\label{Eq: AvgSThetaThetaNH}
%\langle S_{\Theta\Theta}\rangle_{\text{NH}} &=
%-\biggl(p_{o} 
%+
%\frac{\mu}{2 \lambda_{z}^{2}} 
%\frac{R_{o}^{2}}{r_{o}^{2}}
%\biggr) \lambda_{z} \biggl(1 - \frac{\rho^{2}}{r_{o} r_{i}} \biggr)
%+ \frac{\mu}{2}
%\biggl[
%1 -
%\frac{\rho^{4}}{3 r_{o}^{3} r_{i}^{3}} (r_{o}^{2} + r_{o} r_{i} + r_{i}^{2})
%\biggr]
%\notag
%\\
%&\qquad
%-\mu
%\Biggl\{
%\frac{\rho^{2}}{r_{o} r_{i}}
%- \frac{2 \rho}{r_{o} - r_{i}} \ln\frac{(r_{o} - \rho)(r_{i} + \rho)}{(r_{o} + \rho)(r_{i} - \rho)}
%\notag
%\\
%&\qquad\qquad\qquad+
%\frac{1}{r_{o} - r_{i}}
%\biggl[
%\biggl(r_{o} + \frac{\rho^{2}}{r_{o}}\biggr) \ln\frac{R_{o}^{2}}{r_{o}^{2}}
%-
%\biggl(r_{i} + \frac{\rho^{2}}{r_{i}}\biggr) \ln\frac{R_{i}^{2}}{r_{i}^{2}}
%\biggr]
%\Biggr\}.
\end{align}
Again, it is important to keep in mind that, when performing an experiment in which the quantities $\Delta p$, $R_{i}$, $R_{o}$, $r_{i}$, $r_{o}$ and $\lambda_{z}$ are measured and/or controlled, the quantity $\mu$ in Eqs.~\eqref{Eq: AvgqNH} and~\eqref{Eq: AvgsigmathetathetaNH} can be determined using Eqs.~\eqref{Eq: mu bar solution} and~\eqref{Eq: gamma def}.

\paragraph{Mooney-Rivlin material.}
The Cauchy stress for an incompressible homogeneous isotropic Mooney-Rivlin material with undistorted reference configuration is
\begin{equation}
\label{Eq: MR Material}
\bv{\sigma} = -\check{q} \tensor{I} + \mu_{1} (\tensor{B} - \tensor{I}) + \mu_{2} \bigl(\tensor{I} - \tensor{B}^{-1} \bigr),
\end{equation}
where $\check{q}$ is a Lagrange multiplier for the enforcement of incompressibility, and where $\mu_{1}$ and $\mu_{2}$ are constants.  With the substitution $q = \check{q} + \mu_{1} - \mu_{2}$, Eq.~\eqref{Eq: MR Material} becomes
\begin{equation}
\label{Eq: effective MR Material}
\bv{\sigma} = -q \tensor{I} + \mu_{1} \tensor{B} - \mu_{2} \tensor{B}^{-1}.
\end{equation}
In our problem, the $\tensor{B}$ and $\tensor{B}^{-1}$ are given by Eqs.~\eqref{Eq: B} and~\eqref{Eq: B inverse}, respectively. For $\mu_{1} = \mu$ and $\mu_{2} = 0$, Eq.~\eqref{Eq: effective MR Material} yields the neo-Hookean model considered above.  To reduce Eq.~\eqref{Eq: effective MR Material} to a one-parameter model different from the neo-Hookean, we therefore consider a ``reduced Mooney-Rivlin'' model with $\mu_{1} = 0$:
\begin{gather}
\label{Eq: MR special T}
\bv{\sigma} = -\tilde{q} \tensor{I} + 2 \mu_{2} \bv{\varepsilon},
\shortintertext{where}
\tilde{q} = q + \mu_{2}
\quad \text{and} \quad
\tensor{e} = \tfrac{1}{2} \bigl(\tensor{I} - \tensor{B}^{-1}\bigr).
\end{gather}
The tensor $\tensor{e}$ is often referred to as the Eulerian strain tensor (c.f.\ \citealp{Taber_2004_Nonlinear_0}) although other names used are the Hamel (c.f.\ \citealp{Taber_2004_Nonlinear_0}),  Almansi (c.f.\ \citealp{Taber_2004_Nonlinear_0}), and Euler-Almansi (c.f.\ \citealp{Holzapfel-CMBook-2000-1}) strain tensors.  Since $\tensor{B}^{-1}$ is known (see Eq.~\eqref{Eq: B inverse}), the stress $\bv{\sigma}$ in Eq.~\eqref{Eq: effective MR Material} is completely known once the field $q$ is known.  The latter can be shown to be equal to (see Eq.~\eqref{Eq: q RMR final form} in the Appendix)
\begin{equation}
\label{Eq: qMRM solution}
%q = p_{o} + \frac{\Delta p}{\gamma(r_{i},r_{o},R_{i},R_{o},\lambda_{z})} \Biggl[
%\frac{1}{2 \lambda_{z}^{2}} \biggl( \frac{R_{o}^{2}}{r_{o}^{2}} + \frac{R^{2}}{r^{2}} \biggr)
%+
%\frac{1}{\lambda_{z}} \ln\biggl(\frac{r R_{o}}{r_{o} R}\bigg)
%-
%\biggl(
%\frac{r^{2}}{R^{2}} + \frac{1}{\lambda_{z}^{2}} \frac{R^{2}}{r^{2}}
%\biggr)
%\Biggl],
(q)_{\text{RMR}} = p_{o} + \mu_{2} \lambda_{z}^{2} \Biggl[
\frac{1}{2 \lambda_{z}^{2}} \biggl( \frac{R_{o}^{2}}{r_{o}^{2}} + \frac{R^{2}}{r^{2}} \biggr)
+
\frac{1}{\lambda_{z}} \ln\biggl(\frac{r R_{o}}{r_{o} R}\bigg)
-
\biggl(
\frac{r^{2}}{R^{2}} + \frac{1}{\lambda_{z}^{2}} \frac{R^{2}}{r^{2}}
\biggr)
\Biggl],
\end{equation}
where the subscript `$\text{RMR}$' stands for `reduced Mooney-Rivlin'.  Similarly to the neo-Hookean case, $\mu_{2}$ can be expressed in terms of quantities that are measured and/or controlled in an experiment.  Specifically, we have (see Eq.~\eqref{Eq: mu2 solution} in the Appendix)
\begin{equation}
\label{Eq: mubar2 solution}
\mu_{2} = \frac{\Delta p}{\lambda_{z}^{2}\gamma(r_{i},r_{o},R_{i},R_{o},\lambda_{z})},
\end{equation}
where $\gamma$ is given by Eq.~\eqref{Eq: gamma def}.
Substituting $\mu_{1} = 0$ and Eq.~\eqref{Eq: qMRM solution} into Eq.~\eqref{Eq: effective MR Material}, 
%and using Eqs.~\eqref{Eq: F}, \eqref{Eq: FInv}, \eqref{Eq: B inverse}, and~\eqref{Eq: T-S relation} (with $\det \tensor{F} = 1$),
the hoop components of the Cauchy 
%and of the second Piola-Kirchhoff stress tensors are
stress tensor is
\begin{align}
\label{Eq: hoop sigma}
(\sigma_{\theta\theta})_{\text{RMR}} &= -p_{o} - \mu_{2} \lambda_{z}^{2} \Biggl[
\frac{1}{2 \lambda_{z}^{2}} \biggl( \frac{R_{o}^{2}}{r_{o}^{2}} + \frac{R^{2}}{r^{2}} \biggr)
+
\frac{1}{\lambda_{z}} \ln\biggl(\frac{r R_{o}}{r_{o} R}\bigg)
-
\frac{r^{2}}{R^{2}}
\Biggl].
%,
%\\
%\label{Eq: hoop 2ndPK}
%(S_{\Theta\Theta})_{\text{RMR}} &= -p_{o} \frac{R^{2}}{r^{2}} - \lambda_{z}^{2} \mu_{2} \Biggl[
%\frac{1}{2 \lambda_{z}^{2}} \biggl( \frac{R_{o}^{2}}{r_{o}^{2}} + \frac{R^{2}}{r^{2}} \biggr)
%+
%\frac{1}{\lambda_{z}} \ln\biggl(\frac{r R_{o}}{r_{o} R}\bigg)
%-
%\frac{r^{2}}{R^{2}}
%\Biggl] \frac{R^{2}}{r^{2}}
%.
\end{align}
Proceeding as shown for the neo-Hoookean material, the transmural averages of $(q)_{\text{RMR}}$, 
and $(\sigma_{\theta\theta})_{\text{RMR}}$ are
%, and $(S_{\Theta\Theta})_{\text{RMR}}$
%%
\begin{align}
\label{eq: qavg MR}
\langle q \rangle_{\text{RMR}} &= p_{o}
+
\frac{\mu_{2}}{2} \frac{R_{o}^{2}}{r_{o}^{2}}
+
\frac{\mu_{2} \lambda_{z}}{2} \biggl(\frac{\rho^{2}}{r_{o}r_{i}}
-
\frac{2 r_{i}}{r_{o} - r_{i}} \ln\frac{r_{i} R_{o}}{r_{o} R_{i}}
- 3\biggr),
\\
\label{eq: avg sisgmathetathetaRMR}
\langle\sigma_{\theta\theta}\rangle_{\text{RMR}} &= -p_{o}
-
\frac{\mu_{2}}{2} \frac{R_{o}^{2}}{r_{o}^{2}}
+
\frac{\mu_{2} \lambda_{z}}{2} \biggl(\frac{\rho^{2}}{r_{o}r_{i}}
+
\frac{2 r_{i}}{r_{o} - r_{i}} \ln\frac{r_{i} R_{o}}{r_{o} R_{i}}
+ 1\biggr),
%\\
%\label{eq: avg 2ndPKThetaThetaRMR}
%%\langle S_{\Theta\Theta}\rangle_{\text{RMR}} &= -\biggl(p_{o} + \frac{\mu_{2}}{2}  \frac{R_{o}^{2}}{r_{o}^{2}}\biggr)  \frac{R^{2}}{r^{2}}
%%-\frac{\mu_{2}}{2} \frac{R^{4}}{r^{4}}
%%-\lambda_{z} \mu_{2} \frac{R^{2}}{r^{2}} \ln\biggl(\frac{r R_{o}}{r_{o} R}\bigg)
%%+ \lambda_{z}^{2} \mu_{2},
%\langle S_{\Theta\Theta}\rangle_{\text{RMR}} &= 
%-\biggl(p_{o} + \frac{\mu_{2}}{2}  \frac{R_{o}^{2}}{r_{o}^{2}} + \frac{\lambda_{z} \mu_{2}}{2} \ln\frac{r_{o}^{2}}{R_{o}^{2}} \biggr)  \lambda_{z} \biggl(1 - \frac{\rho^{2}}{r_{o} r_{i}}\biggr)
%\notag
%\\
%&\quad+\frac{\lambda_{z}^{2}\mu_{2}}{2} \biggl[1 + 4 \frac{\rho^{2}}{r_{o}r_{i}} -
%\frac{\rho^{4}}{3 r_{o}^{3} r_{i}^{3}} (r_{o}^{2} + r_{o} r_{i} + r_{i}^{2})\biggr]
%\notag
%\\
%&\qquad\;+
%\frac{\lambda_{z}^{2} \mu_{2}}{2(r_{o} - r_{i})}
%\Biggl[
%\biggl(r_{o} + \frac{\rho^{2}}{r_{o}}\biggr) \ln\frac{R_{o}^{2}}{r_{o}^{2}}
%-
%\biggl(r_{i} + \frac{\rho^{2}}{r_{i}}\biggr) \ln\frac{R_{i}^{2}}{r_{i}^{2}}
%- 2\rho\ln\frac{(r_{o} - \rho)(r_{i} + \rho)}{(r_{o} + \rho)(r_{i} - \rho)}
%\Biggr]
%,
\end{align}
where we recall that $\rho = \sqrt{r_{i}^{2} - R_{i}^{2}/\lambda_{z}} = \sqrt{r_{o}^{2} - R_{o}^{2}/\lambda_{z}}$.

\section{Inadequacy of the Laplace law under physiological conditions}
\label{sec: Inadequacy of the Laplace law under physiological conditions}

Using the stress solutions for the two elementary models considered, we now quantitatively assess the accuracy of the Laplace law under normal physiological conditions in the human esophagus.
%We denote by NH and RMR the results obtained using the neo-Hookean  and the reduced Mooney-Rivlin models, respectively, and we present results with lengths nondimensionalized by $r_{i}$ and stresses by $p_{o}$.  For later use, we observe that given Eqs.~\eqref{Eq: B} and~\eqref{Eq: B inverse}, and given the for both the NH and the RMR models, the Cauchy stress has the simple form
%%%
%\begin{equation}
%\label{Eq: Cauchy stress axisymmetric}
%\bv{\sigma} = \sigma_{rr} \, \uv{r} \otimes \uv{r} + \sigma_{\theta\theta}\, \uv{\theta} \otimes \uv{\theta} + \sigma_{zz} \, \uv{z} \otimes \uv{z},
%\end{equation}
%%%
%so that, using Eqs.~\eqref{Eq: T-S relation}, the second Piola-Kirchhoff stress tensor has the form
%%%
%\begin{equation}
%\label{eq: P for T(r)}
%\tensor{S} = \lambda_{z}^{2} \frac{r^{2}}{R^{2}} \sigma_{rr} \, \uv{R} \otimes \uv{R} + \frac{R^{2}}{r^{2}} \sigma_{\theta\theta}\, \uv{\Theta} \otimes \uv{\Theta} + \lambda_{z}^{-2} \sigma_{zz} \, \uv{Z} \otimes \uv{Z}.
%\end{equation}
%%%
%To analyze the validity of the Laplace law for a common physiological application, we choose parameters representative of the human esophagus.
Typical cross-sectional areas and thicknesses of the muscle layers (muscularis propria) of the esophagus have been measured using intraluminal ultrasound \citep{MillerLiu_1995_Correlation_0,NicosiaBrasseur_2002_A-Mathematical_0,Schiffner_2004_Opening_0,MittalPadda_2006_Synchrony_0}.  It is well known that the intrathoracic pressure oscillates with respiration by several mmHg and with a mean value of a couple of mmHg below atmospheric pressure \citep{Christensen_1987_Motor_0}.  A value for the constant $\mu_{2}$ was estimated by \cite{Schiffner_2004_Opening_0} to be approximately \np[mmHg]{60}.  This value was used by \cite{NicosiaBrasseur_2002_A-Mathematical_0} and \cite{GhoshKahrilas_2008_Liquid_0} to model esophageal muscle.  In the current analysis we use the following parameters as representative of the esophagus in the reference state (when $p_{i} = p_{o}$), before inflation by a swallowed bolus of fluid: 
\begin{equation}
\label{Eq: Plot Parameters}
R_{i} = \np[mm]{2.6}, \;
R_{o} = \np[mm]{4.4}, \;
\mu = \mu_{2} = \np[mm\,Hg]{60}, \; 
p_{o} = \np[mm\,Hg]{760}, \;
T/R_{i} = 0.692,
\end{equation}
where $T/R_{i}$ is the muscle layer thickness ratio in the reference state.
Therefore, we evaluate the Laplace law in the range
\begin{equation}
\label{eq: t/ri range}
0 < t/r_{i} < 0.7.
\end{equation}
As alluded to in the Introduction, for the Laplace law to be accurate to within 10\%, it should only be applied when the thickness ratio $t/r_{i}$ is approximately between $0$ and $0.1 \Delta p/p_{o}$. Although one normally swallows liquids and solids as boluses less than \np[ml]{10} in volume, clinical studies are commonly carried out with bolus volumes up to \np[ml]{20}.  Fluoroscopic and endoluminal ultrasound measurements of luminal distension of the esophageal body with bolus volumes up to \np[ml]{20} \citep{Christensen_1987_Motor_0,GhoshKahrilas_2005_The-Mechanical_0,NicosiaBrasseur_2002_A-Mathematical_0,ShiPandolfino_2002_Distinct_0} indicate a lower bound for $t/r_{i}$ of about $0.12$.  Furthermore, it is well known that longitudinal muscle fibers in the esophageal wall contract during bolus transport \citep{DoddsStewart_1973_Movement_0} creating both local and global longitudinal shortening of the esophagus in the range $0.33 < \lambda_{z} < 1.1$ \citep{DaiKorimilli_2006_Muscle_0,MittalPadda_2006_Synchrony_0,NicosiaBrasseur_2002_A-Mathematical_0,ShiPandolfino_2002_Distinct_0}.  Therefore, the range of thickness ratio,
\begin{equation}
\label{Eq: phyiological range def}
0.12 < t/r_{i} < 0.69,
\end{equation}
will be referred to as the \emph{physiological range}.  While we do not focus on longitudinal shortening in this study, to observe how $\lambda_{z}$ may affect the response of the system we present results for $\lambda_{z} = 0.8, 1, 1.1$.

%\subsection{Laplace law and stress estimates}
%\label{subsec: Laplace law and stress estimates}

\paragraph{Stress Estimates.}
In Fig.~\ref{Fig: Deltap over p}
%%%
\begin{figure}[htb]
    \centering
    \includegraphics{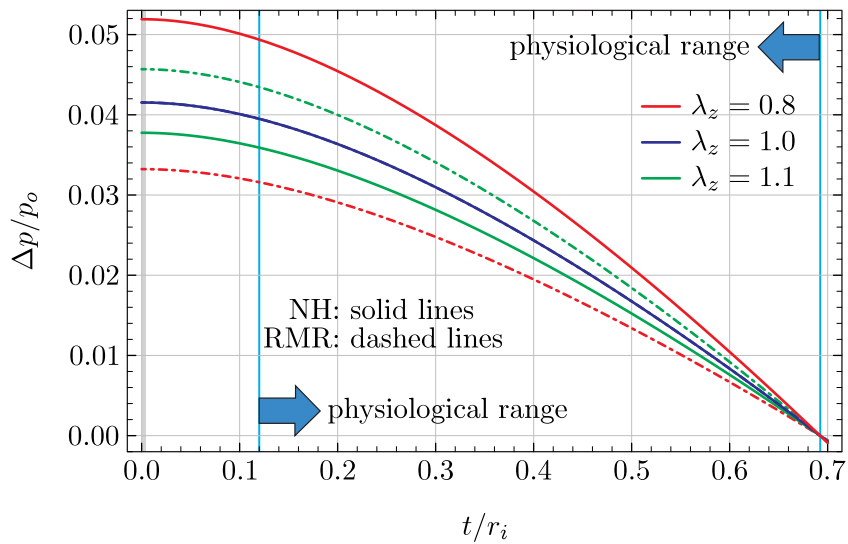}
    \caption{$\Delta p/p_{o}$ vs.\ wall thickness ratio $t/r_{i}$. The behavior of the NH and RMR models is shown via solid and dashed lines, respectively.  Three values of $\lambda_{z}$ are plotted: 0.8 (red), 1 (blue), and 1.1 (green).  The NH and RMR lines for $\lambda_{z} = 1$ coincide.  The physiological range is indicated by vertical blue lines.  The vertical grey strip on the left represents the ``Laplace law'' range defined as $0 < t/r_{i} < 0.1 \Delta p/p_{o}  \approx 0.004$.}
	\label{Fig: Deltap over p}
\end{figure}
%%%
we plot $\Delta p/p_{o}$ as a function of the nondimensional thickness $t/r_{i}$ for the NH and RMR models generated using Eqs.~\eqref{Eq: mu bar solution}, \eqref{Eq: gamma def}, and~\eqref{Eq: mubar2 solution}.  Subject to the constraint in Eq.~\eqref{Eq: Justification for use of LL}, we expect the Laplace law to provide a reasonable estimate of the average hoop stress when  $0 < t/r_{i} < 0.1 \Delta p/p_{o}$.  Figure~\ref{Fig: Deltap over p} shows that, for the NH and RMR models, $\Delta p/p_{o}$ varies between zero and approximately $0.04$, which implies that the Laplace law estimates of the average hoop stress should be reasonably accurate when $0 < t/r_{i} \lessapprox 0.004$, far outside the physiological range.  This is verified in Fig.~\ref{Fig: T over TLL}
%%%
\begin{figure}[htb]
    \centering
    \includegraphics{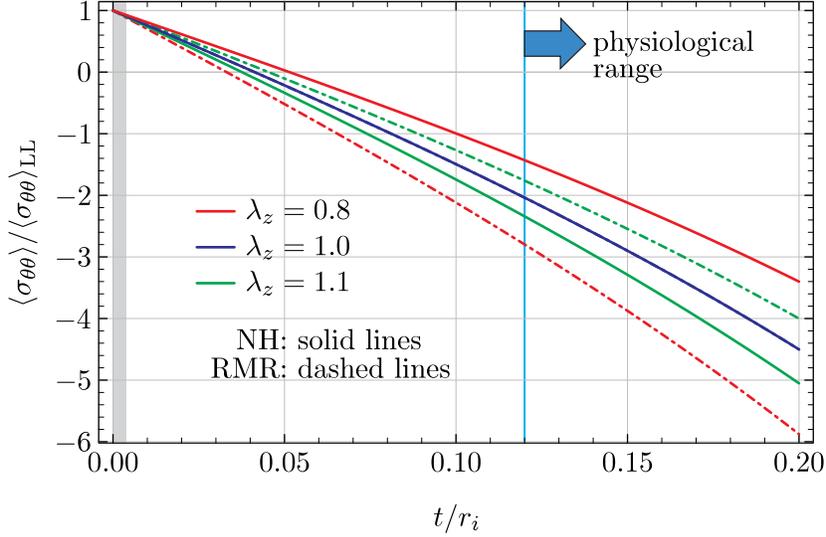}
    \caption{Ratio of exact transmural average stress $\langle \sigma_{\theta\theta} \rangle$ to the corresponding average $\langle \sigma_{\theta\theta} \rangle_{\text{LL}}$ approximated by the Laplace law (Eq.~\eqref{Eq: TLaplacevsTreal}).  The behavior of the NH and RMR models is shown via solid and dashed lines, respectively. Three values of $\lambda_{z}$ are plotted: 0.8 (red), 1 (blue), and 1.1 (green).  The NH and RMR lines for $\lambda_{z} = 1$ coincide.  The physiological range is to the right of the vertical cyan line.  The vertical grey strip on the left represents the ``Laplace law'' range defined as $0 < t/r_{i} < 0.1 \Delta p/p_{o}  \approx 0.004$.}
    \label{Fig: T over TLL}
\end{figure}
%%%
where the ratio of the exact average stress $\langle\sigma_{\theta\theta}\rangle$ (see Eqs.~\eqref{Eq: AvgsigmathetathetaNH} and~\eqref{eq: avg sisgmathetathetaRMR}) to the Laplace law approximation $\langle\sigma_{\theta\theta}\rangle_{\text{LL}}$ in Eq.~\eqref{Eq: Laplace law intro} is plotted against the thickness ratio $t/r_{i}$.  For the NH and RMR models, we have
\begin{equation}
\label{Eq: TLaplacevsTreal}
\frac{\langle \sigma_{\theta\theta} \rangle}{\langle \sigma_{\theta\theta} \rangle_{\text{LL}}} =
%\frac{\langle \sigma_{\theta\theta} \rangle}{\Delta p/(t/r_{i})} = 1 - \frac{p_{o} t}{r_{i} \Delta p} = 
\begin{cases}
1 - \frac{p_{o} t}{r_{i} \gamma \mu} & \text{for the NH model},
\\
1 - \frac{p_{o} t}{r_{i} \lambda_{z}^{2} \gamma \mu_{2}} & \text{for the RMR model}.
\end{cases}
\end{equation}
Figure~\ref{Fig: T over TLL} shows that the the Laplace law provides accurate estimates of the average hoop stress only when $t/r_{i}$ is less than about $0.001$.  Not only are the stress estimates in the physiological range badly in error, they are of the wrong sign.  This is as a result of neglecting $p_{o}$ in the full force balance, Eq.~\eqref{Eq: Laplace law Full}.  That is, the full force balance and the Laplace law differ by the constant $p_{o}$, which is large relative to $\langle\sigma_{\theta\theta}\rangle_{\text{LL}}$, Eq.~\eqref{Eq: Laplace law intro}.  

Since $\langle\sigma_{\theta\theta}\rangle_{\text{LL}}$ and $\langle\sigma_{\theta\theta}\rangle$ differ by the constant $p_{o}$, and since $p_{o}$ must be estimated in any event to evaluate $\Delta p = p_{i} - p_{o}$ from a measurement of $p_{i}$, one could argue that the error in the application of the Laplace law to biological vessels is easily rectified by using the full force balance, Eq.~\eqref{Eq: Laplace law Full}.  However we show next that the characterization of the elastic response of an incompressible material such as muscle using total stress is incorrect, both in principle and in practice.

\section{Pollution of true elastic response by incompressibility}
%\subsection{Estimation of elastic response}
%\label{subsec: Estimation of an elastic response}
%\paragraph{Elastic response.}
In modeling the elastic response of incompressible materials, the total stress should be decomposed as
\begin{equation}
\label{Eq: q sigma-e decomp}
\bv{\sigma} = -q \tensor{I} + \bv{\sigma}^{e},
\end{equation}
where $q$ is the Lagrange multiplier for the enforcement of incompressibility and $\bv{\sigma}^{e}$ is the part of the stress response that is governed by deformation (i.e., the part that is known when an appropriate measure of deformation is known).  To characterize the elastic response of the material, one must characterize the behavior of $\bv{\sigma}^{e}$ rather than $\bv{\sigma}$.  Since $\langle \sigma_{\theta\theta}\rangle$, the transmural average of the hoop component of $\bv{\sigma}$, is much easier to determine experimentally than the $\langle \sigma^{e}_{\theta\theta}\rangle$, we quantify the difference between $\langle \sigma_{\theta\theta}\rangle$ and $\langle \sigma_{\theta\theta}^{e}\rangle$ (by quantifying $\langle q \rangle$), and ask whether the Laplace law estimate $\langle \sigma_{\theta\theta}\rangle_{\text{LL}}$ might approximate $\langle \sigma_{\theta\theta}^{e}\rangle$, since we have established that it cannot be used to model $\langle \sigma_{\theta\theta}\rangle$.  We use the solutions we have obtained for the NH and RMR models to address these issues.

%From the solutions for the field $q$ in Eqs.~\eqref{Eq: sigmabar function of Delta p} and~\eqref{Eq: MR sigmabar Delta p}, we derive the following expressions for the transmural average of $q$ for the NH and RMR models:
%%%
%\begin{align}
%\label{eq: qavg NH}
%\langle q \rangle_{\text{NH}} &= p_{o}
%+
%\frac{\mu}{2 \lambda_{z}^{2}} \frac{R_{o}^{2}}{r_{o}^{2}}
%+ \frac{\mu}{2 \lambda_{z}} 
%\biggl(1 - \frac{\rho^{2}}{r_{o}r_{i}} \biggr)\notag
%\\
%&\quad+
%\frac{\mu}{2 \lambda_{z}(r_{o} - r_{i})} \biggl[ 
%\rho \ln\biggl(
%\frac{(r_{o} - \rho)(r_{i} + \rho)}{(r_{o} + \rho)(r_{i} - \rho)}
%\biggr)
%- 2 r_{i} \ln\biggl(\frac{r_{i} R_{o}}{r_{o}R_{i}}\biggr)
%\biggr],
%\\
%\label{eq: qavg MR}
%\langle q \rangle_{\text{RMR}} &= p_{o}
%+
%\frac{\mu_{2}}{2} \frac{R_{o}^{2}}{r_{o}^{2}}
%+
%\frac{\mu_{2} \lambda_{z}}{2} \biggl(\frac{\rho^{2}}{r_{o}r_{i}}
%-
%\frac{2 r_{i}}{r_{o} - r_{i}} \ln\frac{r_{i} R_{o}}{r_{o} R_{i}}
%- 3\biggr),
%\end{align}
%%%
%where $\rho = \sqrt{r_{i}^{2} - R_{i}^{2}/\lambda_{z}}$.

The expressions in Eqs.~\eqref{Eq: AvgqNH} and~\eqref{eq: qavg MR}, nondimensionalized with respect to $p_{o}$, are plotted in Fig.~\ref{Fig: qavg over po}.
%%%
\begin{figure}[htb]
    \centering
    \includegraphics{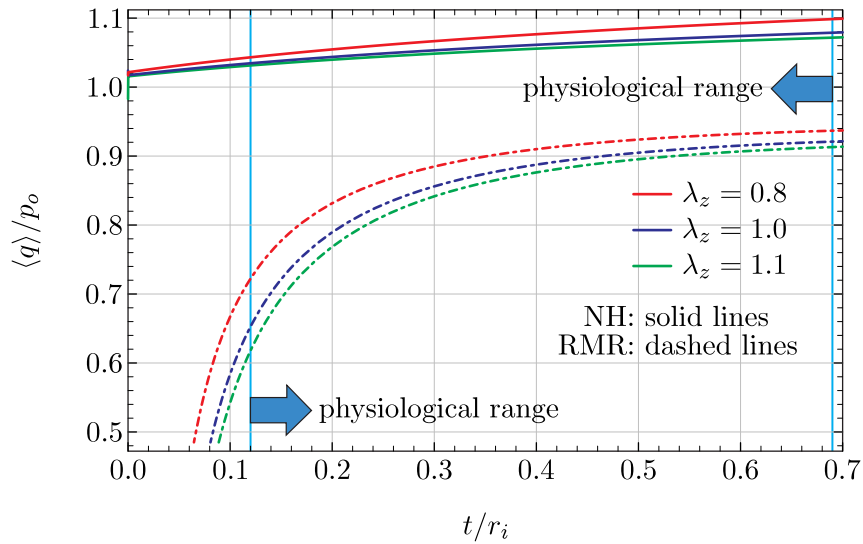}
    \caption{Transmural average of the Lagrange multiplier $q$ vs.\ wall thickness ratio $t/r_{i}$.  The behavior of the NH and RMR models is shown via solid and dashed lines, respectively. Three values of $\lambda_{z}$ are plotted: 0.8 (red), 1 (blue), and 1.1 (green).  The physiological range is between the two vertical cyan lines.}
    \label{Fig: qavg over po}
\end{figure}
%%%
This figure shows that the transmural average of the Lagrange multiplier is strongly material dependent, is comparable to the external pressure, and varies significantly with deformation especially when compared to the variation of the transmural pressure difference over the same range (cf.\ Fig.~\ref{Fig: Deltap over p}).  Thus the effect of incompressibility is a significant fraction of the transmural average of the total hoop stress and should not be ignored.  This is shown explicitly in Fig.~\ref{Fig: q over T},
%%%
\begin{figure}[hbt]
    \centering
    \includegraphics{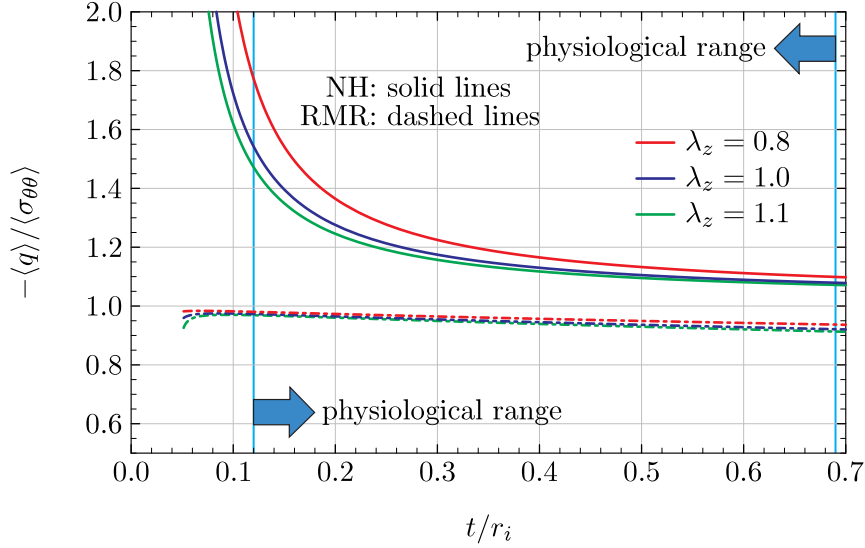}
    \caption{Ratio of the transmural averages of $q$ and $\sigma_{\theta\theta}$ vs.\ wall thickness ratio $t/r_{i}$.  The behavior of the NH and RMR models is shown via solid and dashed lines, respectively.  Three values of $\lambda_{z}$ are plotted: 0.8 (red), 1 (blue), and 1.1 (green).  The physiological range is between the two vertical cyan lines.  All plotted curves approach infinity as $t/r_{i}\to 0$.}
    \label{Fig: q over T}
\end{figure}
%%%
where, over the entire physiological range, the quantity $\langle q \rangle$ is of order or greater than the  average total hoop stress.  In fact, Fig.~\ref{Fig: q over T} indicates that $\langle \sigma_{\theta\theta}\rangle$ is mostly a reflection of $\langle q \rangle$!

We conclude that $\langle \sigma_{\theta\theta}\rangle$ is inappropriate for the determination of a modulus of elasticity.  Consider, for example, the common empirical approach in which the transmural average of the total circumferential stress is plotted against an appropriate strain measure, and a modulus of elasticity is estimated from the slope of the curve, as shown in Fig.~\ref{Fig: P over po vs E}.  Following the approach demonstrated by \citealp{TakedaKassab_2003_Effect_0,TakedaNabae_2004_Oesophageal_0}, in this figure, we plot $\langle S_{\Theta\Theta}\rangle$, the average of the circumferential component of the second Piola-Kirchhoff stress $\tensor{S}$ (see Eqs.~\eqref{Eq: S final app}, \eqref{Eq: S RMR final app}, and Eqs.~\eqref{Aeq: r2R2 and R2r2}--\eqref{Aeq: R4r4 log1 and rRlog avg3} in the Appendix) as a function of $\langle E_{\Theta\Theta}\rangle$, the corresponding average component of the Lagrangian strain $\tensor{E}$, where $\tensor{S}$ and $\tensor{E}$ are defined as follows:\footnote{The tensor $\tensor{E}$ is the work conjugate of $\tensor{S}$.  Also, using Eq.~\eqref{Eq: F}, we have that $E_{\Theta\Theta} = [(r^{2}/R^{2}) - 1]/2$.}
\begin{equation}
\label{Eq: T-S relation and E def}
\tensor{S} = (\det \tensor{F}) \tensor{F}^{-1} \bv{\sigma} \invtrans{\tensor{F}}
\quad \text{and} \quad
\tensor{E} = \tfrac{1}{2} \bigl(\trans{\tensor{F}} \tensor{F} - \tensor{I} \bigr)
\end{equation}
In addition, we show the contributions to $\langle S_{\Theta\Theta} \rangle$ due to the Lagrange multiplier $q$ for the NH and RMR models.  For the RMR model, the contribution due to the Lagrange multiplier is very hard to distinguish from $\langle S_{\Theta\Theta} \rangle$ so the slope in this case would primarily measure a ``modulus of incompressibility,'' rather than a modulus of elasticity.  In the NH case the situation is even worse.  In fact, from Eq.~\eqref{Eq: FInv}, \eqref{Eq: B}, the first of Eqs.~\eqref{Eq: T-S relation and E def}, and \eqref{Eq: Effective stress strain}, we have that $\langle S_{\Theta\Theta} \rangle = -\langle (R^{2}/r^{2}) q \rangle + \mu$, where $-\langle (R^{2}/r^{2}) q \rangle$ is the contribution to $\langle S_{\Theta\Theta} \rangle$ due to the Lagrange multiplier $q$, given by the solid red curve in Fig.~\ref{Fig: q over T}.  Since $\mu$ is constant, for the NH model, the slope of the $\langle S_{\Theta\Theta} \rangle$ curve is \emph{identical} to that of the solid red line; i.e., it is purely a reflection of incompressibility.  The point is that, for incompressible materials, the slope of $\langle S_{\Theta\Theta} \rangle$ vs.\ its conjugate strain has very little to do with the elastic response of the material.  The situation does not improve if other stress measures are used along with other corresponding strain measures.
%%%
\begin{figure}[htb]
    \centering
    \includegraphics{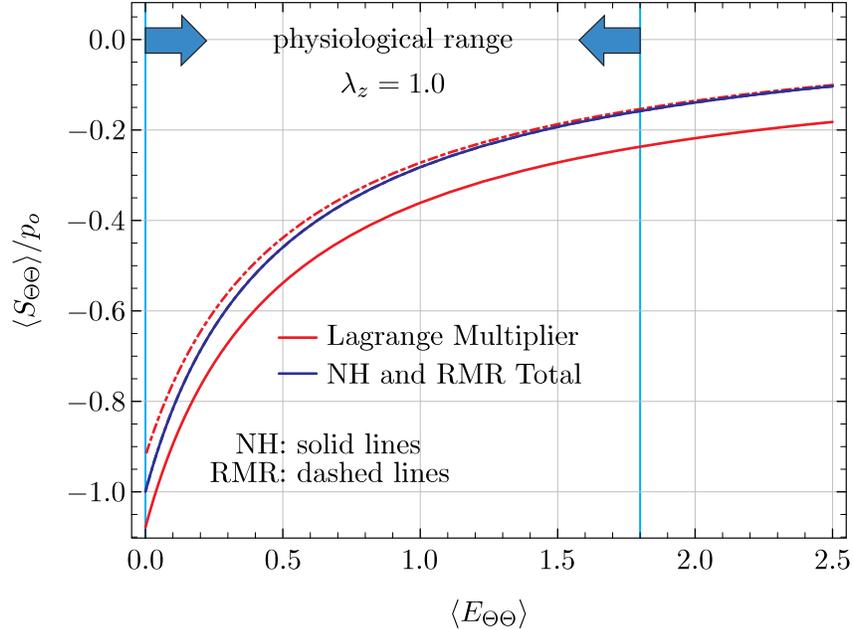}
    \caption{Transmural average of the hoop component of the total second Piola-Kirchhoff stress for $\lambda_{z} = 1$ as a function of the wall averaged hoop component of the Lagrangian strain (blue).  The results for the NH and RMR models are indicated by solid and dashed lines, respectively.  For $\lambda_{z} = 1$ the NH and RMR $\langle S_{\Theta\Theta} \rangle$ curve coincide. The red lines show the contribution to $\langle S_{\Theta\Theta} \rangle$ due to the Lagrange multiplier. The physiological range is delimited by the two vertical cyan lines.}
    \label{Fig: P over po vs E}
\end{figure}
%%%

From this analysis, we conclude  that the elastic behavior of an incompressible material cannot be characterized using the total stress due to the strong influence of the hydrostatic contribution. Is it possible, however, that the Laplace law estimate, which does not approximate the wall averaged total hoop stress, might provide a fortuitous approximation to the slope of the elastic component of the hoop stress $\langle \sigma_{\theta\theta}^{e}\rangle$ when plotted against an appropriate strain measure? For this to be the case, the ratio of $\langle \sigma_{\theta\theta}^{e}\rangle$ to $\langle \sigma_{\theta\theta}\rangle_{\text{LL}}$ must be roughly constant and of order 1 during deformation by inflation. Figure~\ref{Fig: sigma e over sigmaLL}
%%%
\afterpage{\begin{figure}[htb]
    \centering
    \includegraphics{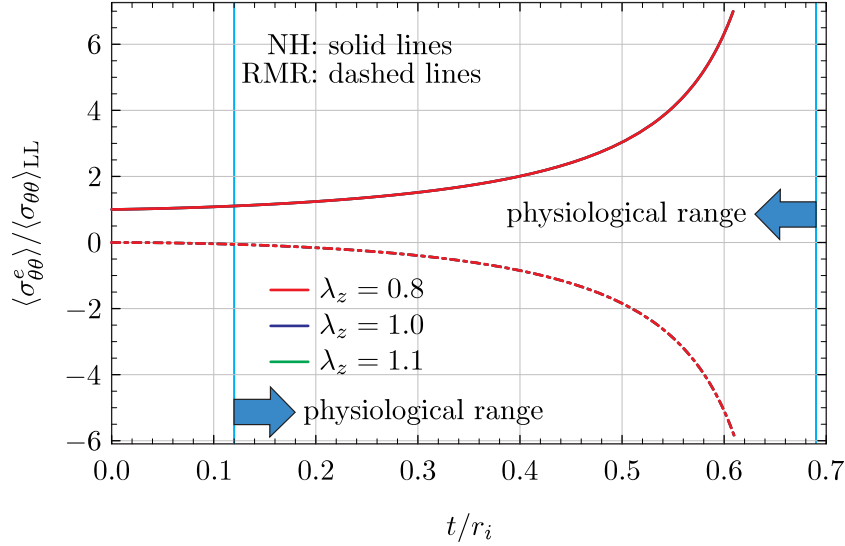}
    \caption{Ratio of the transmural average of the hoop component of $\bv{\sigma}^{e}$ to the correspondent Laplace law approximation of the total stress.  The behavior of the NH and RMR models is shown via solid and dashed lines, respectively. Three values of $\lambda_{z}$ are plotted: 0.8 (red), 1 (blue), and 1.1 (green).  The physiological range is between the two vertical cyan lines.  All of the curves pertaining to the NH and RMR models coincide; only the red curve is visible.}
    \label{Fig: sigma e over sigmaLL}
\end{figure}}
%%%
shows that this is not the case.  Not only does the ratio of $\langle \sigma_{\theta\theta}^{e}\rangle$ to $\langle \sigma_{\theta\theta}\rangle_{\text{LL}}$ strongly vary during inflation of the vessel, its nonlinear variation is strongly material dependent. Thus, another approach is needed to replace the incorrect application of the Laplace law, an approach that ``filters out'' the contribution to total stress due to incompressibility (as well as residual stress) in order to quantify a true elastic modulus, or ``stiffness'' of tubular biological vessels. We present a new practical method to estimate a true modulus of elasticity using the same data obtained with the current approach based on the Laplace law.

\section{A new approach to estimate a true modulus of elasticity for biological tubular vessels}
There is a great need for a method to estimate sensible stiffness measures of tubular muscle, \emph{in vivo} as well as \emph{in vitro}, that can be properly interpreted as moduli of elasticity in order to quantify changes associated with abnormality and disease.  Although, in general, the structure of the luminal walls is not homogeneous or isotropic---in the GI tract, for example, muscle fibers are aligned azimuthally and longitudinally---for such clinical and therapeutic applications it is acceptable to use homogeneous isotropic models if they produce relatively simple measures that represent a true elastic response to deformation. A common current method applies an isotropic model and interprets muscle stiffness from the slope of wall-averaged total hoop stress, estimated from the Laplace law with pressure and geometry measurements, plotted against an appropriate circumferential strain measure. We have shown that this approach does not provide a sensible measure of an elasticity modulus as a quantification of muscle stiffness.

There are two essential difficulties with the common approach. Firstly, the Laplace law does not approximate the true hoop stress for biological tissue and should be avoided in general.  More importantly, to quantify the true elastic response of incompressible materials, the contribution to the stress due to incompressibility must be removed.  If the influence of incompressibility were minor, in principle, the total stress plotted against an appropriate strain might retain a sufficient portion of the elastic response to serve as an approximate measure of stiffness.  However, we have shown that incompressibility can be a dominant influence.

In this section we propose a new approach to estimate muscle stiffness that is not influenced by incompressibility.  The basic observation is that the shear response of an incompressible material is unaffected by incompressibility.  Thus, a true measure of elastic response can be obtained through an \emph{effective elastic shear modulus}. The aim, therefore, is to develop a technique to estimate an effective elastic shear modulus within the standard experimental approach of inflation and extension of tubular organs \emph{in vivo} as well as \emph{in vitro}, i.e., using the same information that is used to estimate muscle stiffness as current methods.

Our approach applies to incompressible tissue and assumes that the reference configuration is undistorted so that the residual stress is hydrostatic (Bowen, 1989) and can be absorbed into the Lagrange multiplier. To develop our approach, we relate simple shear and the inflation and extension test to define an effective shear modulus based on the same experimental method as current approaches.

\subsection{The simple shear test}
We review the simple shear test as a basis for our proposed approach to the determination of an effective shear modulus of tubular vessels from inflation and extension.  The material presented here can be found in several textbooks; we follow \cite{GurtinFried_2010_The-Mechanics_0}.

%Let $\mathcal{I}_{\tensor{B}} = \{I_{1}(\tensor{B}), I_{2}(\tensor{B}), I_{2}(\tensor{B})\}$ denote

The principal invariants of the left Cauchy-Green strain tensor $\tensor{B}$ are 
\begin{equation}
\label{eq: B invariants def}
I_{1} = \trace \tensor{B}, \quad
I_{2}(\tensor{B}) = \tfrac{1}{2}[(\trace\tensor{B})^{2}- \trace(\tensor{B}^{2})],\quad
I_{3}(\tensor{B}) = \det \tensor{B}.
\end{equation}
%%
%The most general form of the constitutive equation of an elastic isotropic material in an undistorted configuration is
%%%
%\begin{equation}
%\label{Eq: CE Cauchy isotropic general}
%\bv{\sigma} = 
%\beta_{0}(\mathcal{I}_{\tensor{B}}) \, \tensor{I}
%+
%\beta_{1}(\mathcal{I}_{\tensor{B}}) \, \tensor{B}
%+
%\beta_{2}(\mathcal{I}_{\tensor{B}}) \, \tensor{B}^{-1},
%\end{equation}
%%%
%where $\beta_{0}(\mathcal{I}_{\tensor{B}})$, $\beta_{1}(\mathcal{I}_{\tensor{B}})$, and $\beta_{3}(\mathcal{I}_{\tensor{B}})$ are three constitutive scalar functions of $\mathcal{I}_{\tensor{B}}$.  For incompressible materials, $I_{3}(\tensor{B}) = 1$ and the elastic stress response of the material is known up to a hydrostatic component dependent on a Lagrange multiplier $q$.  Hence, 
%
The most general form of the constitutive response function for the Cauchy stress of an incompressible homogeneous isotropic material in an undistorted reference configuration is
\begin{equation}
\label{Eq: CE Cauchy isotropic incompressible}
\bv{\sigma} = 
-q \, \tensor{I}
+
\beta_{1}(\tilde{\mathcal{I}}_{\tensor{B}}) \, \tensor{B}
+
\beta_{2}(\tilde{\mathcal{I}}_{\tensor{B}}) \, \tensor{B}^{-1},
\end{equation}
where $\tilde{\mathcal{I}}_{\tensor{B}}$ is the list of only the first two invariants of $\tensor{B}$ (incompressibility demands that $I_{3} = 1$).  Let $(x_{1}, x_{2}, x_{3})$ denote the Cartesian coordinates of a point in the deformed configuration whose Cartesian coordinates in the reference configuration are $(X_{1}, X_{2}, X_{3})$.
%Let $(\uv{1}, \uv{2}, \uv{3})$ be the standard orthonormal basis associated with the coordinate system  $(x_{1}, x_{2}, x_{3})$.
A \emph{simple shear} is a homogeneous deformation of the form
\begin{equation}
\label{Eq: Simple Shear Def}
x_{1} = X_{1} + \zeta X_{2},
\quad
x_{2} = X_{2},
\quad
x_{3} = X_{3},
\end{equation}
where $\zeta = \tan\theta > 0$ is a prescribed scalar parameter measuring the tangent of the shear angle $\theta$. Consider a body that, in its reference configuration, is a rectangular prism with sides $a$, $b$, and $c$ lying along the coordinate axes $X_{1}$, $X_{2}$, and $X_{3}$ respectively (one vertex of the prism coincides with the origin).  For the particular deformation in Eq.~\eqref{Eq: Simple Shear Def}, in the deformed configuration, $(\pi/2) - \theta$ is the angle between sides $a$ and $b$ and the $X_{1}X_{2}$-plane is the shear plane.
%whose geometrical interpretation is illustrated in Fig.~\ref{Fig: Simple Shear}.
%%%
%\begin{figure}[htb]
%    \centering
%    \includegraphics{\FigPath{SimpleShear}}
%    \caption{Geometry of the simple shear test.}
%    \label{Fig: Simple Shear}
%\end{figure}
%%%
Equations~\eqref{Eq: Simple Shear Def} imply that the principal invariants of $\tensor{B}$ for the deformation at hand are
\begin{equation}
\label{SS: Principal Invariants of B}
I_{1}(\tensor{B}) = 3 + \zeta^{2},
\quad
I_{2}(\tensor{B}) = 3 + \zeta^{2},
\quad \text{and} \quad
I_{3} = 1,
\end{equation}
where the last of Eqs.~\eqref{SS: Principal Invariants of B} implies that a simple shear is an isochoric deformation and therefore consistent with incompressibility.  We note that the stretch in the $x_{3}$ direction is equal to 1.  The components of the Cauchy stress of an incompressible homogeneous isotropic material subject to the deformation in Eqs.~\eqref{Eq: Simple Shear Def}, are
\begin{equation}
\label{Eq: Cauchy simple shear}
\begin{bmatrix}
\sigma_{11} & \sigma_{12} & \sigma_{13}
\\
\sigma_{21} & \sigma_{22} & \sigma_{23}
\\
\sigma_{31} & \sigma_{32} & \sigma_{33}
\end{bmatrix}
=
- q
\begin{bmatrix}
1 & 0 & 0
\\
0 & 1 & 0
\\
0 & 0 & 1
\end{bmatrix}
1 + \tilde{\beta}_{1}(\zeta^{2})
\begin{bmatrix}
1 + \zeta^{2} & \zeta & 0
\\
\zeta & 1 & 0
\\
0 & 0  & 1
\end{bmatrix}
+ \tilde{\beta}_{2}(\zeta^{2})
\begin{bmatrix}
1  & - \zeta & 0
\\
- \zeta & 1 + \zeta^{2} & 0
\\
0 & 0  & 1
\end{bmatrix},
\end{equation}
where $\tilde{\beta}_{1}$ and $\tilde{\beta}_{2}$ are the expressions for $\beta_{1}$ and $\beta_{2}$ corresponding to invariants in Eqs.~\eqref{SS: Principal Invariants of B}.  We observe that the shear components of stress are unaffected by the Lagrange multiplier $q$ and that the `$12$' components of the tensors $\tensor{B}$ and $\tensor{B}^{-1}$ are $\zeta$ and $-\zeta$, respectively, that is they are a direct measure of the shear deformation.
%
%It is customary to express the relation linking $\sigma_{12}$ and $B_{12}$ as
%%%
%\begin{equation}
%\label{Eq: SS T12 B12 rel}
%\sigma_{12} = \mu(\zeta^{2}) \zeta,
%\end{equation}
%%%
%where the function $\mu(\zeta^{2})$ is, by definition, the \emph{shear modulus} of the material.  From Eq.~\eqref{Eq: Cauchy simple shear},
%
The \emph{shear modulus} is the ratio between the stress component associated with the shear plane and the corresponding component of $\tensor{B}$.  Referring to Eq.~\eqref{Eq: Simple Shear Def}, recalling that the $X_{1}X_{2}$-plane is the shear plane, and that $B_{12} = \zeta$, the shear modulus is given by
%%%
%\begin{equation}
%\label{Eq: SS mu def}
%\mu(\zeta^{2}) = \tilde{\beta}_{1}(\zeta^{2}) - \tilde{\beta}_{2}(\zeta^{2}).
%\end{equation}
%%%
%In a general sense, we can view the shear modulus as being defined by the ratio
%%
\begin{equation}
\label{Eq: SS T12 B12 rel B}
\mu = \frac{\sigma_{12}}{B_{12}} = \tilde{\beta}_{1}(\zeta^{2}) - \tilde{\beta}_{2}(\zeta^{2}).
\end{equation}
To distinguish this shear modulus determined via a simple shear test from the effective shear modulus introduced later, we will refer to $\mu$ as the ``true'' shear modulus.

\subsection{Shear response via the inflation and extension test}
\label{subsection: new method}
For isotropic materials, and in fact even for orthotropic materials if the directions of orthotropy are appropriately aligned, the principal directions of strain coincide with the corresponding principal directions of the Cauchy stress.  In the extension and inflation test, the principal directions of deformation are parallel to the coordinate lines of the coordinate system (Fig.~\ref{fig: coordinate systems}).  Using elementary tensor transformations, we deduce that, at each point, the maximum shear strain occurs along lines in the principal planes that bisect the coordinate directions.

Consider a cross section perpendicular to the axis of the cylinder, as shown in Fig.~\ref{fig: xi ets systems},
%%%
\begin{figure}[htb]
    \centering
    \includegraphics{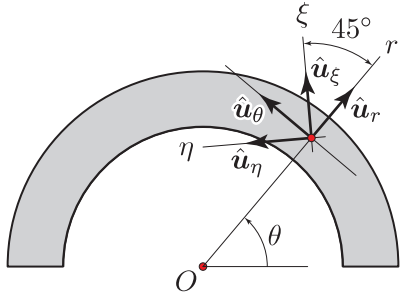}
    \caption{Directions of maximum shear.}
    \label{fig: xi ets systems}
\end{figure}
%%%
where the ($\theta$-dependent) lines $\xi$ and $\eta$ are defined such that $\xi$ forms a $45^{\circ}$ angle with the $r$ axis.  The orthonormal unit vectors $\uv{\xi}$ and $\uv{\eta}$ in the $\xi$ and $\eta$ directions can be expressed in terms of the unit vectors $\uv{r}$ and $\uv{\theta}$ as follows:
\begin{equation}
\label{eq: uxi and ueta}
\uv{\xi} = \frac{1}{\sqrt{2}} (\uv{r} + \uv{\theta})
\quad \text{and} \quad
\uv{\eta} = \frac{1}{\sqrt{2}} (-\uv{r} + \uv{\theta}).
\end{equation}
Since $\tensor{B}$ and $\bv{\sigma}$ have the forms $\tensor{B} = B_{rr} \, \uv{r} \otimes \uv{r} + B_{\theta\theta} \, \uv{\theta} \otimes \uv{\theta} + B_{zz} \, \uv{z} \otimes \uv{z}$ and $\bv{\sigma} = \sigma_{rr} \, \uv{r} \otimes \uv{r} + \sigma_{\theta\theta} \, \uv{\theta} \otimes \uv{\theta} + \sigma_{zz} \, \uv{z} \otimes \uv{z}$, the $\xi\eta$ (shear) components of these tensors are
\begin{equation}
\label{eq: shear comps of B and T}
B_{\xi\eta} = \uv{\xi} \cdot \tensor{B} \uv{\eta} = \tfrac{1}{2} (B_{\theta\theta} - B_{rr})
\quad \text{and} \quad
\sigma_{\xi\eta} = \uv{\xi} \cdot \bv{\sigma} \uv{\eta} = \tfrac{1}{2} (\sigma_{\theta\theta} - \sigma_{rr}).
\end{equation}
Note that the shear stress $\sigma_{\xi\eta}$ is completely unaffected by incompressibility and it can be argued that it is also unaffected by the presence of a residual stress state.\footnote{This is certainly the case for isotropic materials in an undistorted configuration.  It is also the case for orthotropic materials in an undistorted configuration whose directions of orthotropy are parallel to the principal directions of strain in the inflation and extension test.}

We now introduce the logarithmic average through the wall of the vessel, which we denote by $\llbracket \bullet \rrbracket$ and define as follows:
%Another important consideration about $\sigma_{\xi\eta}$ is that it is straightforward to compute its logarithmic average experimentally in an extension and inflation test, where the logarithmic average\footnote{The introduction of the logarithmic average is motivated by the observation that, in the present context, integrals in $\d{r}/r$ arise naturally from the equilibrium equations (see the second of Eqs.~\eqref{Eq: Trr balance law}).} of a generic quantity $\phi$, which we denote by $\llbracket \phi \rrbracket$, is defined as
%%
\begin{equation}
\label{eq: log avg def}
\llbracket \bullet \rrbracket := \frac{1}{\ln(r_{o}/r_{i})} \int_{r_{i}}^{r_{o}} \bullet \, \frac{\d{r}}{r}.
\end{equation}
Using the above definition, along with the last of Eqs.~\eqref{eq: shear comps of B and T} and the equilibrium equation in Eq.~\eqref{Eq: NHs div T = 0}, which is valid for any constitutive model for which the principal directions of stress are parallel to the principal directions of strain in the inflation and extension test, we have
\begin{equation}
\label{eq: log avg Txieta}
%\llbracket \sigma_{\xi\eta} \rrbracket = \frac{1}{2 \ln(r_{o}/r_{i})} \int_{r_{i}}^{r_{o}} \frac{\sigma_{\theta\theta} - \sigma_{rr}}{r} \, \d{r}
%\quad \Rightarrow \quad
\llbracket \sigma_{\xi\eta} \rrbracket = \frac{\Delta p}{2 \ln(r_{o}/r_{i})},
\end{equation}
where we have used the pressure boundary conditions on the inner and outer surfaces of the cylinder (with the reference state taken as the no load state).

The stress estimate in Eq.~\eqref{eq: log avg Txieta} is as easy to obtain experimentally as that of current methods based on Eq.~\eqref{Eq: Laplace law intro} or its correct form in Eq.~\eqref{Eq: Laplace law Full}.   However, the estimate in Eq.~\eqref{eq: log avg Txieta} is completely unaffected by incompressibility.  In addition, and most importantly, using Eq.~\eqref{eq: log avg Txieta} we can obtain an effective (i.e., not pointwise) yet rigorous measure of a shear modulus of the material, again immune from incompressibility effects.  Mimicking the definition of shear modulus from the analysis of the simple shear test, and denoting the vessel's \emph{effective} shear modulus by $\mu_{\text{eff}}$, we define $\mu_{\text{eff}}$ as follows:
\begin{equation}
\label{eq: mueff def}
\mu_{\text{eff}} := \frac{\llbracket \sigma_{\xi\eta} \rrbracket}{\llbracket B_{\xi\eta} \rrbracket}.
\end{equation}
That is, we define effective the (wall averaged) shear modulus as the ratio of the Cauchy shear stress to the corresponding Cauchy-Green shear strain, where both are log-averaged over the esophageal wall with the lumen represented as an effective cylinder.  Using Eq.~\eqref{Eq: B} and the first of Eqs.~\eqref{eq: shear comps of B and T} we have that, in an inflation and extension test, $B_{\xi\eta} = (1/2)[(r^{2}/R^{2})-R^{2}/(\lambda_{z}^{2} r^{2})]$, so that
\begin{equation}
\label{eq: Bxieta}
%\llbracket B_{\xi\eta} \rrbracket = \frac{1}{2 \ln(r_{o}/r_{i})}
%\int_{r_{i}}^{r_{o}} \frac{B_{\theta\theta} - B_{rr}}{r} \, \d{r}
%\quad \Rightarrow \quad
\llbracket B_{\xi\eta} \rrbracket =
\frac{1}{2 \ln(r_{o}/r_{i})}
\int_{r_{i}}^{r_{o}} \biggl(\frac{r^{2}}{R^{2}} - \frac{R^{2}}{\lambda_{z}^{2} r^{2}}\biggr) \, \frac{\d{r}}{r}.
\end{equation}
%%
%where we have used Eq.~\eqref{Eq: B}.
As it turns out, the integral in Eq.~\eqref{eq: Bxieta} coincides with the function $\gamma(r_{i},r_{o},R_{i},R_{o},\lambda_{z})$ in Eq.~\eqref{Eq: gamma def}.  Therefore,
\begin{equation}
\label{eq: Bxieta final}
\llbracket B_{\xi\eta} \rrbracket = \frac{\gamma}{2 \ln(r_{o}/r_{i})},
\end{equation}
so that, substituting Eqs.~\eqref{eq: log avg Txieta} and~\eqref{eq: Bxieta final} into Eq.~\eqref{eq: mueff def}, and using the definition in Eq.~\eqref{Eq: gamma def},
\begin{equation}
\label{eq: mueff full}
%\llbracket B_{\xi\eta} \rrbracket = \frac{\gamma}{2 \ln(r_{o}/r_{i})}
%\quad \text{and} \quad
\mu_{\text{eff}} = \frac{\Delta p}{\gamma} = \Delta p \biggl[\frac{1}{2 \lambda_{z}^{2}} \biggl( \frac{R_{o}^{2}}{r_{o}^{2}} - \frac{R_{i}^{2}}{r_{i}^{2}} \biggr) + \frac{1}{\lambda_{z}} \ln\biggl( \frac{r_{i} R_{o}}{r_{o} R_{i}} \biggr)\biggr]^{-1},
\end{equation}
where, for the class of deformations being considered here and from Eq.~\eqref{Eq: continuity}, $\lambda_{z} = (R_{o}^{2} - R_{i}^{2})/(r_{o}^{2} - r_{i}^{2})$ is the ratio of cross sectional wall areas in the reference state relative to the deformed state.   For isotropic incompressible tissue, this ratio quantifies the stretch in longitudinal direction of the esophageal wall relative to the (no-load) reference state,\footnote{The same could be said for orthotropic materials such that the longitudinal axis of the esophagus is also a direction of material symmetry} and $\lambda_{z} < 1$ implies longitudinal shortening.

The effective shear modulus $\mu_{\text{eff}}$ can be used in experiments as a rigorous measure of the elastic shear response of any incompressible isotropic material.
%Despite the limitation concerning isotropy, the class of models in question is ample and includes models appropriate for the study of the esophagus and other biological vessels.
Since the formula for $\mu_{\text{eff}}$ is explicitly given, its computation does not require the determination of a slope from a graph of stress vs.\ strain.  As such, Eq.~\eqref{eq: mueff full} offers a direct estimation of the elastic response of the material that is more straightforward to obtain than that of current approaches.  Clearly, $\llbracket \sigma_{\xi\eta} \rrbracket$ can always be plotted against $\llbracket B_{\xi\eta} \rrbracket$ if desired.  However, because the stiffness measure $\mu_{\text{eff}}$ is the ratio of stress to strain, it must be interpreted as a secant modulus rather than as a tangent modulus, which is the slope of a stress vs.\ strain curve.

\begin{remark}[Comparison with the simple shear test]
The quantity $\mu_{\text{eff}}$ is defined as the ratio of two average quantities. Therefore, it pertains to a structure as a whole rather than being the outcome of a homogeneous deformation (like the simple shear test), which can be said to apply pointwise.  With this in mind, to compare  $\mu_{\text{eff}}$ with the true shear modulus, referring to the discussion following Eq.~\eqref{Eq: Cauchy simple shear}, we recall that, in the simple shear test, the components $B_{12}$ and $B^{-1}_{12}$ are such that $B_{12} = - B^{-1}_{12}$.  Furthermore, the principal stretch in the direction perpendicular to the shear plane is equal to 1.  A similar situation occurs in the inflation and extension test when $\lambda_{z} = 1$, that is, $\llbracket B_{\xi\eta} \rrbracket = -\llbracket B^{-1}_{\xi\eta} \rrbracket$. Therefore, we will compare the true shear modulus of a variety of models with the correspondent $\mu_{\text{eff}}$ only for $\lambda_{z} = 1$.  However, the definition of $\mu_{\text{eff}}$ is not constrained by $\lambda_{z} = 1$ in that it always represents a rigorous measure of the elastic shear response of the material.
\end{remark}

\subsection{Comparison between $\mu_{\text{eff}}$ and the true shear modulus for various models}

Using the definition of $\mu_{\text{eff}}$ and Eq.~\eqref{Eq: CE Cauchy isotropic incompressible}, we derive expressions for $\mu_{\text{eff}}$ for some common constitutive models.  To facilitate these derivations, we first derive expressions for $\tensor{B}$ and $\tensor{B}^{-1}$ in terms of the principal stretches $\lambda_{r} = R/(\lambda_{z}r)$ and $\lambda_{z}$ ($\lambda_{\theta} = (\lambda_{z}\lambda_{r})^{-1}$).  From Eq.~\eqref{Eq: B},
\begin{align}
\label{Eq: B ps}
\tensor{B} &= \lambda_{r}^{2} \, \uv{r} \otimes \uv{r} + \frac{1}{\lambda_{z}^{2}\lambda_{r}^{2}} \, \uv{\theta} \otimes \uv{\theta} + \lambda_{z}^{2} \, \uv{z} \otimes \uv{z}
\shortintertext{and}
\label{Eq: BInv ps}
\tensor{B}^{-1} &= \frac{1}{\lambda_{r}^{2}}\, \uv{r} \otimes \uv{r} + \lambda_{z}^{2} \lambda_{r}^{2} \, \uv{\theta} \otimes \uv{\theta} + \frac{1}{\lambda_{z}^{2}} \, \uv{z} \otimes \uv{z}.
\end{align}
The corresponding expressions for $B_{\xi\eta}$ and $B^{-1}_{\xi\eta}$ are
\begin{equation}
\label{eq: Bxieta and Bxietainv ps}
B_{\xi\eta} = \frac{1 - \lambda_{z}^{2} \lambda_{r}^{4}}{2 \lambda_{z}^{2} \lambda_{r}^{2}}
\quad \text{and} \quad
B^{-1}_{\xi\eta} = -\frac{1 - \lambda_{z}^{2} \lambda_{r}^{4}}{2 \lambda_{r}^{2}},
\end{equation}
which imply
\begin{equation}
\label{eq: Bxieta and Binvxieta}
B_{\xi\eta}^{-1} = -\lambda_{z}^{2} B_{\xi\eta}.
\end{equation}
Using the definition of $\mu_{\text{eff}}$ and Eq.~\eqref{Eq: CE Cauchy isotropic incompressible},
\begin{equation}
\label{eq: mueff general}
\mu_{\text{eff}} = \frac{\llbracket \beta_{1}(\tilde{\mathcal{I}}_{\tensor{B}}) B_{\xi\eta}\rrbracket + 
\llbracket \beta_{2}(\tilde{\mathcal{I}}_{\tensor{B}}) B^{-1}_{\xi\eta}\rrbracket
}{\llbracket B_{\xi\eta}\rrbracket}.
\end{equation}

\paragraph{Neo-Hookean and Mooney-Rivlin Models.}
Since $\beta_{1} = \mu$ and $\beta_{2} = 0$ for the neo-Hookean model, and $\beta_{1} = \mu_{1}$ and $\beta_{2} = -\mu_{2}$ for the (full) Mooney-Rivlin model, Eq.~\eqref{eq: mueff general} yields
\begin{equation}
\label{eq: mueff NH and MR}
(\mu_{\text{eff}})_{\text{NH}} = \mu
\quad \text{and} \quad
(\mu_{\text{eff}})_{\text{MR}} = \mu_{1} + \mu_{2}
\frac{\lambda_{z}^{2}(\lambda_{r_{o}}^{2} - \lambda_{r_{i}}^{2}) + 2 \lambda_{z} \ln(\lambda_{r_{o}}/\lambda_{r_{i}})}{(\lambda_{r_{o}}^{2} - \lambda_{r_{i}}^{2}) + (2/ \lambda_{z}) \ln(\lambda_{r_{o}}/\lambda_{r_{i}})}.
% \frac{\llbracket B_{\xi\eta}^{-1} \rrbracket}{\llbracket B_{\xi\eta}\rrbracket}.
\end{equation}
Since $\mu$ is the true shear modulus of a neo-Hookean material, the first of Eqs.~\eqref{eq: mueff NH and MR} implies that $(\mu_{\text{eff}})_{\text{NH}}$ is exactly equal to the true shear modulus.  To compare $(\mu_{\text{eff}})_{\text{MR}}$ to the true shear modulus of the MR model we let $\lambda_{z} = 1$ in the second of Eqs.~\eqref{eq: mueff NH and MR}.  This gives
\begin{equation}
\label{eq: mueff MR lambda = 1}
(\mu_{\text{eff}})_{\text{MR}}\big|_{\lambda_{z}=1} = \mu_{1} + \mu_{2},
\end{equation}
which, again, is the true shear modulus for the (full) Mooney-Rivlin model.

\paragraph{Gent Model.}
We now consider the Gent model (see, e.g., \citealp{GurtinFried_2010_The-Mechanics_0}):
\begin{equation}
\label{eq: TGent}
\bv{\sigma} = -q \tensor{I} + 2 c_{1} \biggl( 1 - \frac{I_{1} - 3}{I_{1}^{m}} \biggr)^{\!-1} \tensor{B} - 2 c_{2} I_{2}^{-1} \tensor{B}^{-1},
\end{equation}
where $I_{1}$ and $I_{2}$ are defined in Eqs.~\eqref{eq: B invariants def}, and where $c_{1} \geq 0$, $c_{2} \geq 0$, and $I_{1}^{m} > 3$ are constants.  The true shear modulus of the Gent material is \citep{GurtinFried_2010_The-Mechanics_0}
\begin{equation}
\label{eq: muGent}
\mu = 2 c_{1} \biggl( 1 - \frac{\zeta^{2}}{I_{1}^{m}} \biggr)^{\!-1} + \frac{2 c_{2}}{3 + \zeta^{2}}.
\end{equation}
The functions $\beta_{1}$ and $\beta_{2}$ for the Gent model in an inflation and extension test are
\begin{equation}
\label{eq: b1 and b2 Gent}
\beta_{1} = \frac{2 c_{1} I_{1}^{m} \lambda_{z}^{2} \lambda_{r}^{2}}{(I_{1}^{m} + 3 - \lambda_{z}^{2})\lambda_{z}^{2} \lambda_{r}^{2} - \lambda_{z}^{2} \lambda_{r}^{4} - 1}
\quad \text{and} \quad
\beta_{2} = - \frac{2 c_{2} \lambda_{z}^{2} \lambda_{r}^{2}}{\lambda_{z}^{4} \lambda_{r}^{4} + \lambda_{z}^{2} + \lambda_{r}^{2}}.
\end{equation}
Therefore,
%letting
%%
%\begin{equation}
%\label{eq: Gent k def}
%k = I_{1}^{m} + 3 - \lambda_{z}^{2},
%\end{equation}
%%
the $\mu_{\text{eff}}$ for the Gent model is
\begin{multline}
\label{eq: mueff Gent}
(\mu_{\text{eff}})_{\text{Gent}} =
c_{1} I_{1}^{m}
\frac{%%
\frac{2(2 + k \lambda_{z})}{\sqrt{4 - k^{2} \lambda_{z}^{2}}}
\biggl[
\tan^{-1}\biggl(\lambda_{z}\frac{k-2\lambda_{r_{o}}^{2}}{\sqrt{4 - k^{2} \lambda_{z}^{2}}}\biggr)
-
\tan^{-1}\biggl(\lambda_{z}\frac{k-2\lambda_{r_{i}}^{2}}{\sqrt{4 - k^{2} \lambda_{z}^{2}}}\biggr)
\biggr]
-
\ln\frac{ 1 - k \lambda_{z}^{2} \lambda_{r_{o}}^{2} + \lambda_{z}^{2} \lambda_{r_{o}}^{4}}{ 1 - k \lambda_{z}^{2} \lambda_{r_{i}}^{2} + \lambda_{z}^{2} \lambda_{r_{i}}^{4}}
}
{%%
\bigl[\lambda_{r_{o}}^{2} - \lambda_{r_{i}}^{2} + (2/\lambda_{z}) \ln(\lambda_{r_{o}}/\lambda_{r_{i}})\bigr]
}
\\
+ c_{2}
\frac{%%
\frac{2(2 \lambda_{z}^{3} - 1)}{\sqrt{4 \lambda_{z}^{6} - 1}}
\biggl[
\tan^{-1}\biggl(
\frac{2 \lambda_{z}^{4} \lambda_{r_{o}}^{2} + 1}{\sqrt{4 \lambda_{z}^{6} - 1}}\biggr)
-
\tan^{-1}\biggl(
\frac{2 \lambda_{z}^{4} \lambda_{r_{i}}^{2} + 1}{\sqrt{4 \lambda_{z}^{6} - 1}}\biggr)
\biggr]
+
\ln\frac{\lambda_{z}^{4}\lambda_{r_{o}}^{4} + \lambda_{r_{o}}^{2} + \lambda_{z}^{2}}{\lambda_{z}^{4}\lambda_{r_{i}}^{4} + \lambda_{r_{i}}^{2} + \lambda_{z}^{2}}
}
{%%
\bigl[\lambda_{r_{o}}^{2} - \lambda_{r_{i}}^{2} + (2/\lambda_{z}) \ln(\lambda_{r_{o}}/\lambda_{r_{i}})\bigr]
},
\end{multline}
where $k = I_{1}^{m} + 3 - \lambda_{z}^{2}$.  To compare $(\mu_{\text{eff}})_{\text{Gent}}$ with the true shear modulus in Eq.~\eqref{eq: muGent}, let $\lambda_{z} = 1$ in Eq.~\eqref{eq: mueff Gent} to obtain
\begin{align}
\label{eq: mueff gent small}
(\mu_{\text{eff}})_{\text{Gent}}\big|_{\lambda_{z} = 1}
&=
\Biggl\{
c_{1} I_{1}^{m}
\frac{2(4 + I_{1}^{m})}{\sqrt{4 - (2 + I_{1}^{m})^{2}}}
\biggl[
\tan^{-1}\biggl(\frac{2 + I_{1}^{m} -2\lambda_{r_{o}}^{2}}{\sqrt{4 - (2 + I_{1}^{m})^{2}}}\biggr)
-
\tan^{-1}\biggl(\frac{2 + I_{1}^{m}-2\lambda_{r_{i}}^{2}}{\sqrt{4 - (2 + I_{1}^{m})^{2}}}\biggr)
\biggr]
\notag
\\
&\qquad-
c_{1} I_{1}^{m}
\ln\frac{ 1 - (2 + I_{1}^{m}) \lambda_{r_{o}}^{2} + \lambda_{r_{o}}^{4}}{ 1 - (2 + I_{1}^{m}) \lambda_{r_{i}}^{2} + \lambda_{r_{i}}^{4}}
\notag
\\
&\qquad+
c_{2}
\frac{2}{\sqrt{3}}
\biggl[
\tan^{-1}\biggl(
\frac{2 \lambda_{r_{o}}^{2} + 1}{\sqrt{3}}\biggr)
-
\tan^{-1}\biggl(
\frac{2 \lambda_{r_{i}}^{2} + 1}{\sqrt{3}}\biggr)
\biggr]
\notag
\\
&\qquad+
c_{2} \ln\frac{\lambda_{r_{o}}^{4} + \lambda_{r_{o}}^{2} + 1}{\lambda_{r_{i}}^{4} + \lambda_{r_{i}}^{2} + 1}
\Biggr\}
\bigl[\lambda_{r_{o}}^{2} - \lambda_{r_{i}}^{2} + 2 \ln(\lambda_{r_{o}}/\lambda_{r_{i}})\bigr]^{-1}.
\end{align}
We compare Eqs.~\eqref{eq: muGent} and~\eqref{eq: mueff gent small} in Fig.~\ref{fig: GentModelShear},
%%%
\afterpage{\begin{figure}[htb]
    \centering
    \includegraphics{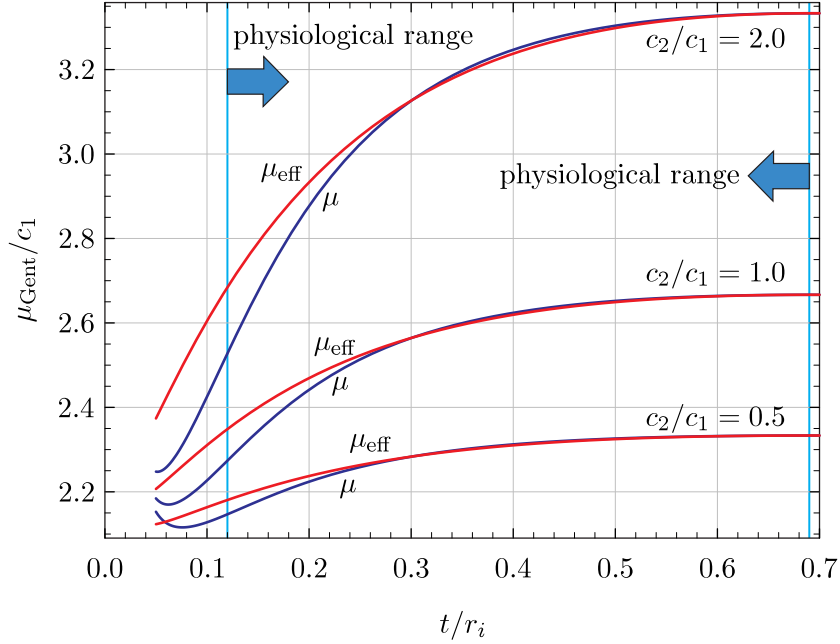}
    \caption{Comparison between the nondimensional true shear modulus and the effective shear modulus for the Gent Model.  The nondimensionalization was carried out relative to the $c_{1}$ parameter appearing in the Gent model.  The solid blue lines correspond to Eq.~\eqref{eq: muGent} with $\zeta = \llbracket B_{\xi\eta}\rrbracket$ and are denoted by $\mu$. The solid red lines correspond to Eq.~\eqref{eq: mueff gent small} and are denoted by $\mu_{\text{eff}}$.  The two cyan lines delimit the physiological range.}
    \label{fig: GentModelShear}
\end{figure}}
%%%
where we have set $\zeta$ in Eq.~\eqref{eq: muGent} equal to $\llbracket B_{\xi\eta} \rrbracket\big|_{\lambda_{z} = 1}$ in the first of Eqs.~\eqref{eq: mueff full}, i.e.,
\begin{equation}
\label{eq: muGent comparison id}
\zeta = \frac{1}{4 \ln(r_{o}/r_{i})} \bigl[
\lambda_{r_{o}}^{2} - \lambda_{r_{i}}^{2} + 2 \ln(\lambda_{r_{o}}/\lambda_{r_{i}})
\bigl].
\end{equation}
We non-dimensionalized the expressions for the shear moduli with $c_{1}$, and have chosen the following values of the constitutive parameters:
\begin{equation}
\label{eq: muGent comparison Parameters}
\frac{c_{2}}{c_{1}} = \tfrac{1}{2}, 1, 2
\quad \text{and} \quad
I_{1}^{m} = 500.
\end{equation}
In the Gent model, the quantity $I_{1}^{m}$ serves as an upper bound of the quantity $I_{1}$.  If $I_{1}^{m}$ is too small, it is not possible to explore the behavior of the material in the thin wall limit as we have done earlier.  Recalling that the physiological range in the esophagus had $0.12 < t/r_{i} < 0.69$, we can see that the effective shear modulus tracks the true shear modulus well in the physiological range.  The difference between the two shear moduli is well below $10\%$ even around $t/r_{i} = 0.1$.  To put things in perspective, the value $t/r_{i} = 0.1$ corresponds to $\llbracket B_{\xi\eta} \rrbracket = 2.664$, which, in turn, corresponds to a shear angle of $69.43^{\circ}$.  This level of shear is extreme for the biological systems of interest.

\paragraph{Fung Model.}
We now consider an isotropic incompressible Fung-like model of the type \citep{Fung_1981_Biomechanics:_0}
\begin{equation}
\label{eq: TFung}
\bv{\sigma} = -q \tensor{I} + b e^{a(I_{1} - 3)} \tensor{B},
\end{equation}
where $a$ and $b$ are constants and $I_{1} = \trace\tensor{B}$.  The true shear modulus for this model is
\begin{equation}
\label{eq: mu Fung}
\mu = b e^{a \zeta^{2}}.
\end{equation}
In an inflation and extension test, the functions $\beta_{1}$ and $\beta_{2}$ for this model are
\begin{equation}
\label{eq: beta1 and beta2 Fung}
\beta_{1} = b e^{a(\lambda_{r}^{2} + \lambda_{r}^{-2} \lambda_{z}^{-2} + \lambda_{z}^{2} - 3)}
\quad \text{and} \quad
\beta_{2} = 0.
\end{equation}
As simple as the expression for $\beta_{1}$ may seem, we were unable to obtain a closed-form expression for the $\mu_{\text{eff}}$ of the Fung model.  Therefore, our results are based on a numerical evaluation of $\mu_{\text{eff}}$ carried out using \emph{Mathematica}~8 \citep{Wolfram-Research2010Mathematica0}.  In Fig.~\ref{fig: FungModelShear},
%%%%
\afterpage{\begin{figure}[htb]
    \centering
    \includegraphics{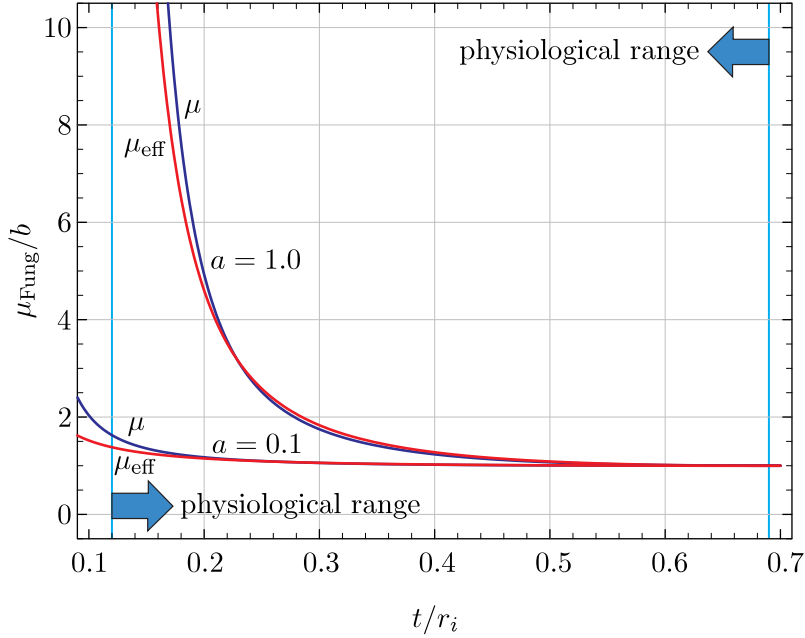}
    \caption{Comparison between the nondimensional shear modulus and the effective shear modulus for the Fung model.  The nondimensionalization was carried out relative to the $b$ parameter appearing in the Fung model.  The solid blue lines correspond to Eq.~\eqref{eq: mu Fung} with $\zeta = \llbracket B_{\xi\eta}\rrbracket$ and are denoted by $\mu$. The solid red lines correspond to Eq.~\eqref{eq: mueff general} for the Fung model and are denoted by $\mu_{\text{eff}}$.  The two cyan lines delimit the physiological range.}
    \label{fig: FungModelShear}
\end{figure}}
%%%%
we plot $\mu/b$ from Eq.~\eqref{eq: mu Fung} along with $\mu_{\text{eff}}$ obtained by substituting Eqs.~\eqref{eq: beta1 and beta2 Fung} into Eq.~\eqref{eq: mueff def} for the case with $\lambda_{z} = 1$.  We considered two cases with $a = 0.1$ and $1$.  As was done in Fig.~\ref{fig: GentModelShear} for the Gent model, for the purpose of comparison, we let $\zeta = \llbracket B_{\xi\eta} \rrbracket\big|_{|\lambda_{z} = 1}$ in Eq.~\eqref{eq: mu Fung}.  Figure~\ref{fig: FungModelShear} shows that the approximation of $\mu$ by $\mu_{\text{eff}}$ is very good over the entire physiological range.  Not surprisingly, the quality of the approximation decreases for very small values of $t/r_{i}$ corresponding to extremely large the shear strains that are generally outside the physiological ranges of interest.

\section{Example: a re-evaluation of the standard method from the literature}
\label{section: Reevaluation of the standard method}
We have shown that the slope of total hoop stress plotted against strain does not approximate a true material stiffness in principle and that the Laplace law does not approximate hoop stress for typical tubular vessels in the human body in practice. To illustrate our points, we here re-evaluate data given in \cite{TakedaKassab_2003_Effect_0} using our proposed method to estimate a true effective shear modulus with the same data collected for the Laplace law method. We used \citeauthor{TakedaKassab_2003_Effect_0}\ to illustrate our method primarily because they have presented data in plots and statistical measures that make possible a re-evaluation and comparison using our proposed method.

We refer to the Laplace-law-based approach applied by \citeauthor{TakedaKassab_2003_Effect_0}\ as the ``standard approach.'' The aim of the experiments by \citeauthor{TakedaKassab_2003_Effect_0}\ was to determine the effects of atropine, a smooth muscle relaxant, on esophageal wall stiffness, where ``stiffness'' was quantified by applying the standard approach to estimate what was thought to be an extensional modulus of elasticity (see Section~\ref{section: introduction}), as described here below.

In \cite{TakedaKassab_2003_Effect_0}, intraluminal pressure was measured in the esophagus with a manometric catheter at the same location as an intraluminal ultrasound probe that imaged the cross section of the esophageal wall, $\approx \np[cm]{7.5}$ above the lower esophageal sphincter (high pressure zone). Approximating the pressure external to the esophageal wall as atmospheric and estimating the thickness of the esophageal wall from the ultrasound images by mapping the cross sectional areas into circles with equivalent areas, the average total circumferential stress of the esophageal wall was estimated using the Laplace equation, namely Eq.~\eqref{Eq: Laplace law intro}. To estimate wall stiffness, \citeauthor{TakedaKassab_2003_Effect_0}\ collected data with systematic distension of the esophageal lumen by progressively filling a high-compliance bag within the lumen with water. The bag was attached to the manometric catheter so that bag pressure and volume was measured simultaneously during distension. Using the changes in intraluminal pressure from manometry and changes in cross sectional geometry estimated from ultrasound images, wall hoop stress was quantified using the Laplace law and circumferential strain was quantified, relative to the resting state, as relative change in effective circumferential dimension of the esophageal lumen. \citeauthor{TakedaKassab_2003_Effect_0}\ then used slopes of stress against strain plots as an elastic modulus to compare esophageal wall properties before vs.\ after administering atropine.

\citeauthor{TakedaKassab_2003_Effect_0}\ collected their distension data using two different protocols. In one protocol, the bag was filled in steps by systematically increasing bag volume (isovolumic distension) from $5$ to $\np[ml]{20}$. In the second protocol the bag was filled by systemically increasing bag pressure (isobaric distension) from $10$ to $\np[mmHg]{60}$. The level of distension and the time over which the lumen was distended in each step were different for the two protocols. The rate of distension and the time delay before data were collected, potentially affecting the generation of active tone, were not given. Nevertheless, one expects similar changes in wall properties in response to atropine, independent of protocol.

For $8$--$10$ asymptomatic subjects the following quantifications were estimated at each distension by \citeauthor{TakedaKassab_2003_Effect_0}: intraluminal pressure, lumen cross sectional area, outer wall circumference, and approximate wall thickness. The ratio of outer circumference in the distended vs.\ resting state was computed at each distention and was used to convert the Laplace law hoop stress to the corresponding component of the second Piola-Kirchhoff stress and to calculate the corresponding component of the Lagrangian strain. Plots of (aforementioned components of) the second Piola-Kirchhoff stress against Lagrangian strain were fit with straight lines over the ranges of applied distension, the slopes of which were interpreted as moduli of elasticity.

A primary conclusion from \citeauthor{TakedaKassab_2003_Effect_0}\ was stated in the abstract:
\begin{quote}
The YoungÕs modulus, which is the slope of a linear stress-strain relationship, was significantly higher after atropine in the isovolumic study but not in the isobaric study.Ó
\end{quote}
Specifically, \citeauthor{TakedaKassab_2003_Effect_0}\ estimated the YoungÕs modulus to be \np[mmHg]{37} vs.\ \np[mmHg]{107} before vs.\ after atropine with isovolumic distension, while with isobaric distension they estimated \np[mmHg]{102} before atropine vs.\ \np[mmHg]{107} after atropine was applied, a change that was not statistically significant. This result lead to the following conclusion (see the discussion section of \citealp{TakedaKassab_2003_Effect_0}):
\begin{quote}
These results suggest that atropine behaves as if it stiffens the esophageal body in an isovolumic but not an isobaric study,
\end{quote}
followed the obvious question
\begin{quote}
Why does atropine behave in different ways depending on the modality of esophageal distension?
\end{quote}
\citeauthor{TakedaKassab_2003_Effect_0}\ go on to argue that their result implies a protocol-dependent active response to atropine overlying a protocol-independent passive response.

However, we have argued in this paper that \emph{total} stress response to strain is seriously polluted by incompressibility in the hydrostatic component, that the tangent to the curve of a total principal stress component against its corresponding strain component therefore does not provide a true elastic modulus, and that the standard approach is fundamentally erroneous, so that a true modulus requires filtering out the hydrostatic component from total stress. In addition, we have shown that Laplace law estimates for total stress are seriously in error for the esophagus and other thick-walled elastic biological vessels, and that the Laplace law stress cannot be used as a surrogate for the strain-dependent component of total stress. We re-evaluate the \citeauthor{TakedaKassab_2003_Effect_0}\ result with our approach based on an effective shear stress that is unpolluted by incompressibility and avoids the Laplace law. With this re-evaluation, we also point out differences in application between our proposed method and the standard approach.

\subsection{Application of the new method to the data by \cite{TakedaKassab_2003_Effect_0}}

Whereas the standard approach uses the slope of total hoop stress against strain to define an elasticity modulus, our approach estimates an effective shear modulus directly with Eq.~\eqref{eq: mueff full}.  Like the standard approach, we replace the true lumen cross section by an effective circular cross section with equivalent cross sectional areas. The outer and inner radii of the effective lumen are determined in the reference configuration $(R_{i},R_{o})$ and under distension $(r_{i},r_{o})$. To apply Eq.~\eqref{eq: mueff full} one first quantifies the reference configuration, assumed to be the no-load state (i.e., the state with zero transmural pressure difference, $\Delta p$). Then, with an inflation experiment (isobaric or isovolumic), the transmural pressure difference $\Delta p$ is quantified together with inner and outer radius $r_{i}$ and $r_{o}$ of the effective lumen at each distension. Since manometry measures only intraluminal pressure, to estimate $\Delta p$ the external pressure is assumed to be close to atmospheric. Finally, longitudinal stretch in the wall material $\lambda_{z}$ must be estimated from the second of Eqs.~\eqref{Eq: continuity}, which is a statement of incompressibility during distension from $(R_{i}, R_{o})$ to $(r_{i},r_{o})$. With these quantifications, one applies Eq.~\eqref{eq: mueff full} to estimate effective shear modulus.

Note that the data required to estimate $\lambda_{z}$ and effective shear modulus are the same as the data required for the standard approach: inner and outer radii needed in Eq.~\eqref{eq: mueff full} are also needed to quantify hoop stretch in the form of $r_{i}$ and wall thickness $(r_{o} - r_{i})$. With our new method, however, particular care should be taken to quantify the geometry of the reference state as accurately as possible.

Whereas the same information needed in the standard approach is also applied in our new method, our new method requires the user to quantify explicitly the local longitudinal stretch $\lambda_{z}$, a physiologically important quantity worth studying in its own right (see, e.g., \citealp{NicosiaBrasseur_2001_Local_0,ShiPandolfino_2002_Distinct_0,MittalPadda_2006_Synchrony_0}), and to incorporate the quantification of longitudinal stretch into our effective shear modulus. The fact that the traditional approach does not use longitudinal stretch to estimate stiffness is a manifestation of not removing the important confounding effects of incompressibility from total stress to produce the stress component that represents elastic response. 

To extract the information we needed for the application of our method, i.e., to apply Eq.~\eqref{eq: mueff full}, we digitized data presented in plot form in \cite{TakedaKassab_2003_Effect_0}. It should be recognized, however, that had the experiment been designed to directly apply Eq.~\eqref{eq: mueff full}, it would have been developed and analyzed somewhat differently, and we would have had available the raw pre-analyzed data. To obtain the necessary quantifications, we digitized the ensemble-averaged results presented in Figs.~1, 2, and~3 of \citeauthor{TakedaKassab_2003_Effect_0} In the isovolumic case, we quantified pressure differences from Fig.~1A, assuming that \citeauthor{TakedaKassab_2003_Effect_0}\ approximated intrathoracic pressure external to the esophageal wall as atmospheric, as is typical. From Fig.~1B we quantified average intraluminal cross sectional area (CSA).

Although the esophageal lumen cross section is not circular in general (cf.\ \citealp{NicosiaBrasseur_2001_Local_0}), to estimate \emph{effective} properties of the lumen wall, we placed the CSA within a circular cross section to quantify inner radius so that $r_{i}^{2} = \pi/\text{CSA}$. We then used the data from Fig.~3A to obtain measurements of the hoop component of the second Piola-Kirchhoff stress $\sigma$ (according to the notation in \citealp{TakedaKassab_2003_Effect_0}) and the corresponding Lagrangian strain $\epsilon$ (Eq.~(3) in \citealp{TakedaKassab_2003_Effect_0}). From $\epsilon$, we obtained the corresponding circumferential stretch ratio, $\lambda_{\theta}$ using $\epsilon = \tfrac{1}{2} (\lambda_{\theta}^{2} - 1)$, as given in Eq.~(3) of \citeauthor{TakedaKassab_2003_Effect_0} From the Laplace law relationship between the hoop component of the second Piola-Kirchoff stress and wall thickness used by \citeauthor{TakedaKassab_2003_Effect_0}, specifically their Eq.~(5), we were able to extract the wall thickness $h$ and, from that, $r_{o} = r_{i} + h$. To estimate the outer radius in the reference state $R_{o}$, we used the averages for outer circumference supplied by \citeauthor{TakedaKassab_2003_Effect_0} (see p.~624) before and after the administration of atropine. To obtain $R_{i}$, we applied a linear regression to the thickness vs.\ pressure data and extrapolated to $\Delta p = 0$. Combining these results with $R_{o}$ yielded the desired inner radii values in the reference configuration.  A similar strategy was used to analyze the data in the isobaric case.

Since the analysis developed in this paper was based on the averages presented in \cite{TakedaKassab_2003_Effect_0}, we carried out a sensitivity analysis within the range of values provided by \citeauthor{TakedaKassab_2003_Effect_0}\ for the outer circumference in the rest state. We found that the essential results are insensitive to the uncertainly in circumference and the conclusions are unchanged.

\subsection{Analysis of the Effective modulus with the new method}
As discussed above, to quantify the effective shear modulus via Eq.~\eqref{eq: mueff full}, it is necessary to estimate the local longitudinal shortening of the muscle layers given by $\lambda_{z}$ in Eq.~\eqref{Eq: continuity} evaluated at the outer surface of the lumen. The quantity $\lambda_{z}$ measures the local longitudinal dimension of the lumen relative to the longitudinal dimension in the reference (no-load) state. Relative shortening (lengthening) of that axial location in the lumen is quantified by reduction (increase) in $\lambda_{z}$ from $1$. Using the data from the plots in \cite{TakedaKassab_2003_Effect_0} as described above, the variation of $\lambda_{z}$ due to isobaric (pressure control) and isovolumic (volume control) distension of the water-filled bag, as described above, is shown in Fig.~\ref{fig: LambdaPlot}, in which $\lambda_{z}$ is plotted as a function of intraluminal pressure relative to atmospheric pressure.
\begin{figure}[htb]
    \centering
    \includegraphics{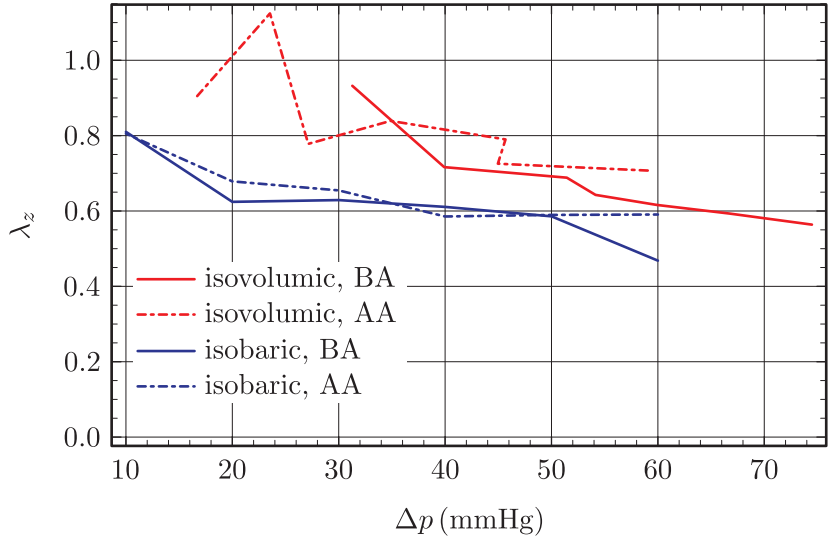}
    \caption{Local longitudinal shortening ($\lambda_{z}$) of the esophageal wall plotted against transmural pressure difference $\Delta p$ for isovolumic and isobaric lumen distension before (BA) and after (AA) atropine was administered, as per protocols and data presented in \cite{TakedaKassab_2003_Effect_0}.}
    \label{fig: LambdaPlot}
\end{figure}
Figure~\ref{fig: LambdaPlot} indicates that the esophageal lumen at the measurement location did shorten longitudinally in response to distension, indicating the activation of longitudinal muscle fibers. The degree of shortening increased with increasing distension at the same rate with both protocols, both before and after atropine was administered. The isobaric protocol, in which pressure in the bag was increased stepwise, suggests that the level of longitudinal muscle shortening is about the same before and after atropine, while the isovolumic distension result is less clear since the relationship between pressure and distension was significantly altered by the smooth-muscle-relaxing effects of atropine \citep{TakedaKassab_2003_Effect_0}. This is indicated in Fig.~\ref{fig: LambdaPlot} by the shift in the post-atropine curve to the left relative to the pre-atropine curve. The results also seem to suggest that the level of longitudinal shortening at the same intraluminal pressure is protocol dependent, with greater longitudinal shortening under pressure control than volume control. However, the relative shifts in the curves change when $\lambda_{z}$ is plotted against distention ($r_{i}$) rather than pressure, so one cannot draw general conclusions from the comparison of magnitude of longitudinal shortening between protocols and before vs.\ after atropine. Nevertheless, distension stimulates longitudinal shortening in all cases and the level of shortening increases with increasing distention at about the same rate for both protocols and before vs.\ after administering atropine. We conclude, therefore, that the stimulation of longitudinal shortening at the same rate before vs.\ after administering atropine is a general result, independent of protocol.

The effective shear modulus of the esophageal wall, $\mu_{\text{eff}}$, is plotted in Fig.~\ref{fig: EffectiveMuab}%
{\afterpage{%
\begin{figure}[htb]
    \centering
    \includegraphics{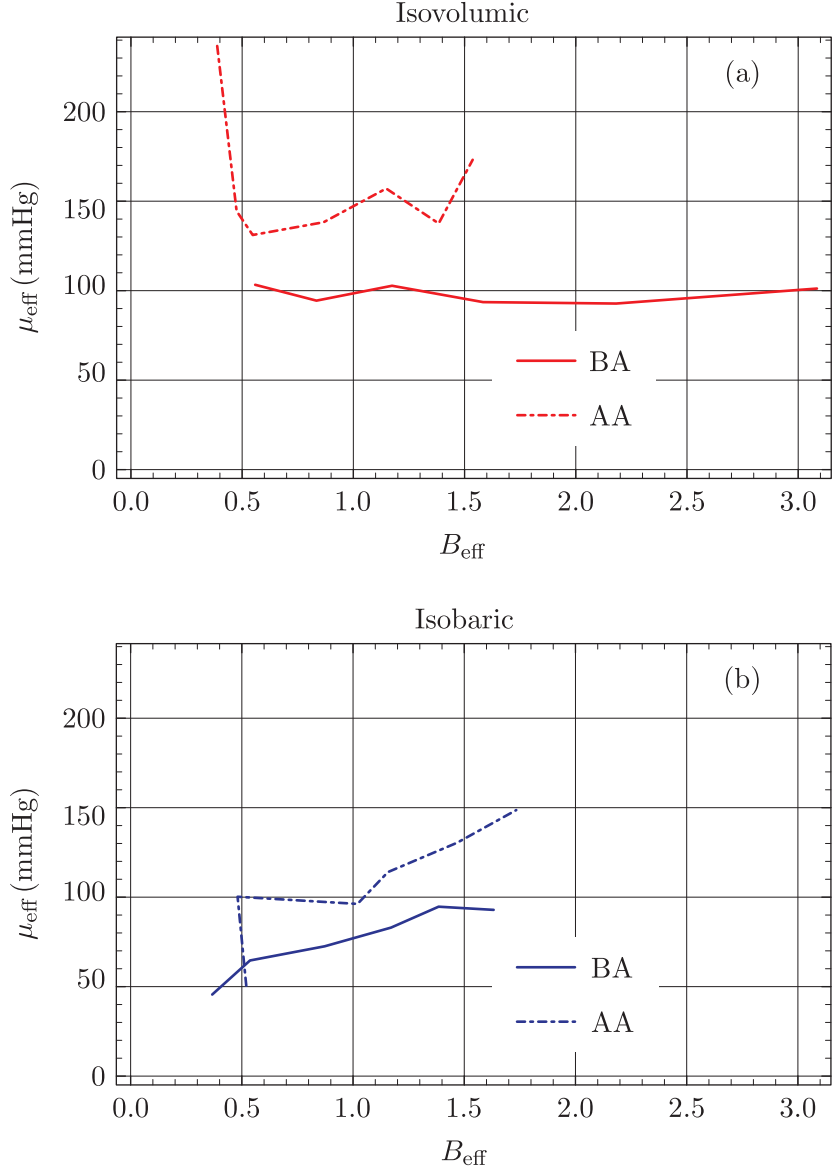}
    \caption{Effective shear modulus of the esophageal wall $\mu_{\text{eff}}$ (Eq.~\eqref{eq: mueff full}) plotted against the component $B_{\text{eff}} = \llbracket B_{\xi\eta} \rrbracket$ of the effective left Cauchy-Green strain in Eq.~\eqref{eq: Bxieta final} for (a) isovolumic and (b) isobaric lumen distension before (BA) and after (AA) atropine was administered.}
    \label{fig: EffectiveMuab}
\end{figure}}
against the effective shear component of the left Cauchy-Green strain, $B_{\text{eff}} = \llbracket B_{\xi\eta} \rrbracket$, using the data in \cite{TakedaKassab_2003_Effect_0} and the results for longitudinal shortening parameter in Fig.~\ref{fig: LambdaPlot}. Comparisons are made before vs.\ after the administration of atropine for the isobaric and isovolumic protocols in which bag pressure or bag volume were controlled during distension, and comparing effective wall stiffness. Note that the two after atropine (AA) results become unstable in the limit of low strain, when $B_{\text{eff}} \approx 0.5$. This is because in Eq.~\eqref{eq: mueff full} both the numerator and the denominator approach zero in the reference state limit, since $B_{\text{eff}}$ and $\Delta p$ are both zero in the reference state by definition. To maintain finite $\mu_\text{eff}$ as $B_{\text{eff}} \to 0$, the denominator must approach zero at the same rate as the numerator; however the ratio is sensitive to imprecision in the measured data and truncation error in the no-load limit. The unstable range must therefore be excluded in the analysis of $\mu_{\text{eff}}$, and in Fig.~\ref{fig: EffectiveMuab} we only analyze the range $B_{\text{eff}} > 0.5$.

The standard method used by \citeauthor{TakedaKassab_2003_Effect_0}, where an apparent modulus is deduced by plotting the Laplace law approximation of total hoop stress against strain, led to inconsistent results depending on which protocol was used to distend the esophageal wall: the apparent modulus increased after atropine when the esophagus was distended by incremental increases in bag volume, but did not change when the esophagus was distended by increasing bag pressure. The apparent result that the effect of atropine on esophageal wall stiffness is protocol-dependent led  \citeauthor{TakedaKassab_2003_Effect_0}\ to suggest that the tonic response of esophageal muscle is different depending on how the esophagus is distended. Atropine is a smooth muscle relaxant; it is unclear why the change in muscle stiffness from unrelaxed to relaxed state would depend on the method used to distend muscle.

Figure~\ref{fig: EffectiveMuab}, however, shows that the inconsistency disappears when the standard approach is replaced by the application of the effective shear modulus developed in Section~\ref{subsection: new method}: Fig.~\ref{fig: EffectiveMuab} shows that the effective shear modulus increased after administering atropine in both the isovolumnic and isobaric protocols. The change in muscle stiffness with atropine is independent of the protocol used to distend the esophageal wall. We conclude that the resistance to distension is greater when the esophageal smooth muscle is in an atropine-induced relaxed state than in the normal state in which the response is a summation of passive elastic resistance and changes in active tone stimulated by stretch. This conclusion, based on a true measure of elastic modulus, is a general one, independent of protocol, resolving the previously inconsistent result. The resolution of inconsistency, we argue, is a consequence of removing the two essential errors in the standard method: (1) the pollution of the hydrostatic component of the total stress response to strain by incompressibility, and (2) the inaccurate representation of both total and elastic stress by the Laplace law.

Having generalized the essential result that esophageal muscle is stiffer after atropine, we can note potential application-dependent differences in the results of Fig.~\ref{fig: EffectiveMuab}. In particular, although \citeauthor{TakedaKassab_2003_Effect_0}\ using the standard method concluded that the elastic modulus was independent of the magnitude of distension in all cases, Fig.~\ref{fig: EffectiveMuab} suggests that this may be the case for the isovolumic protocol, but that with the isobaric protocol the esophageal wall appeared to stiffen with increasing distension, both before and after atropine was administered. Additionally, \citeauthor{TakedaKassab_2003_Effect_0}\ report apparent elastic moduli $\approx \text{$37$--$83$}\,\text{mmHg}$ before-after atropine with isovolumic distension and $\approx \np[mmHg]{105}$ with isobaric distension, while our shear modulus estimates are $\approx \text{$50$--$90$}\,\text{mmHg}$ before atropine and $\approx\text{$100$--$150$}\,\text{mmHg}$ after atropine. Whereas the standard method indicated an increase in elastic modulus of $\approx \np[mmHg]{46}$ after atropine with the isobaric protocol and no change with isovolumic distension, we estimate a change in effective shear modulus of $\text{$30$--$50$}\,\text{mmHg}$ depending on distension level and, perhaps slightly, distension protocol. However without access to the raw data and a more careful statistical analysis, one should not draw firm conclusions from these details.

Our overall conclusions are (1) that the standard method does not provide a proper measure for elastic modulus due to pollution of the hydrostatic component of the total stress due to incompressibility and serious inaccuracies in the Laplace law approximation of total hoop stress, and (2) that our proposed method based on shear stress alleviates the problem by isolating the elastic component of stress and providing a method for estimating stress without use of the Laplace law.

\section{Summary and Conclusions}
In this paper we discuss the applicability of a current common approach to estimate total average hoop stress across the wall of biological vessels, and the practice of estimating material stiffness as a modulus of elasticity, from a plot of total hoop stress against circumferential strain. We refer to this combination as ``the standard method.'' We have shown that both elements of the standard method, the use of total stress estimate an elastic modulus when plotted against strain and the application of the Laplace law approximation to estimate total hoop stress, are inapplicable for biological vessels which are relatively thick and are composed of incompressible tissue such as muscle.

The Laplace law provides an incorrect estimate of the transmural average of the total hoop stress because the crucial assumption, that the relative thickness of the vessel wall is very small in comparison to the relative transmural pressure difference, is rarely met in biological vessels. In comparison to bubbles, for which the Laplace law was originally intended, the Laplace law does not apply to biological vessels as a consequence of the relatively thick nature of the vessel walls, and the deformability of biological tissue in response to relatively low transmural pressure differences. In contrast, the Laplace law is generally applicable to finite-thickness metal tubes when they are supporting very large relative pressure differences. However, in any application for which an estimate of the transmural average of total hoop stress is sought in a human or animal vessels, we argue in this paper that the Laplace law should be avoided.

The Laplace law is a straightforward reduction of a simple equilibrium force balance where the pressure external to the vessel appears only within the transmural pressure difference and the independent effect of external pressure is neglected. However, it is common to observe the Laplace law as a force balance reduction with the external pressure set to zero in all terms. This approximation is not valid in biological vessels and the common practice of setting the external pressure to zero in the Laplace law should be avoided.
The Laplace law estimate of the average total hoop stress is easily corrected by simply including the external pressure in the force balance and not making the ``thin wall approximation.'' The common practice of estimating the elastic modulus of the wall from the slope of the average total hoop stress plotted against an appropriate strain measure should remain valid since, from Eq.~\eqref{Eq: Laplace law Full}, the thin wall approximation simply shifts the average stress response by the amount $p_{o}$, thus leaving unaffected the slope of this response function. However there is a more serious issue in this practice that invalidates the interpretation of slope as a response to the elastic deformation: the incompressibility of muscle contaminates the total stress with a contribution that is largely independent of the elastic response of the material. To interpret the slope of total stress against strain as elastic modulus, one assumes that the contribution due to incompressibility is either independent of deformation or sufficiently small relative to total stress. We have shown that incompressibility can be a major contributor to total hoop stress (Fig.~\ref{Fig: q over T}). Thus, in general, a plot of transmural average of the total hoop stress against strain cannot be used with confidence to estimate a modulus of elasticity, even in an approximate fashion.

We have proposed an alternative approach to overcome the difficulties inherent to the use of the average total hoop stress to estimate a modulus of elasticity. The proposed approach uses the same information typically gathered during an inflation and extension test to derive a mechanistically rigorous measure of the shear response of the material. It is a practical method that, like the current approach, can be applied \emph{in vivo} as well as \emph{in vitro}. The rationale for this approach is that the shear response for isotropic materials and, in some cases, for orthotropic materials, is completely immune from the effects of incompressibility.

The proposed approach provides a measure of an effective shear modulus of the material. We have shown via the analysis of several common models for incompressible isotropic materials that the proposed approach delivers very good estimates of the shear modulus of the material especially over the range of deformations that are of interest in the characterization of biological tissues.

We further re-analyzed data from \cite{TakedaKassab_2003_Effect_0} presented in their plots and statistics. \citeauthor{TakedaKassab_2003_Effect_0}\ used the standard method (deduction of a modulus from the slope of total hoop stress plotted against strain and using the Laplace law to approximate total hoop stress) to estimate an apparent modulus of elasticity with two different protocols to distend the esophageal lumen, one in which the volume of a water filled back in increased in controlled incremental fashion, and one in which water is added to the bag to increase the pressure in the bag in controlled increments. They found that the change in stiffness after administering atropine, a smooth-muscle relaxant, was different depending on the protocol used to distend the esophagus: the esophageal wall apparently stiffened after atropine in one protocol but not the other. We showed that this inconsistency is resolved when our new method for estimating a true shear modulus is applied: the elastic shear modulus is observed to increase after atropine is administered under esophageal distension with either protocol.

We argue that the previous ``standard'' method to estimate a modulus of elasticity of biological vessel walls based on total hoop stress and the Laplace law should be abandoned. We further suggest that our method is a useful replacement for the standard method. Our proposed new approach provides a true measure of elastic modulus without the polluting effects of incompressibility in methods based on total hoop stress and without application of the Laplace law which is highly inaccurate for biological vessels. A further major advantage of our approach is that the effective shear modulus of the material can be measured directly from experimentally obtained quantities. This is different from current approaches, whereby a modulus is inferred from the slope of an experimentally determined stress-strain curve. In this sense, the proposed approach offers a direct accurate measure of the elastic shear response of incompressible isotropic materials.

%%%%%%%%%%%%%%%%%%%%%%%%%%%%%%%%%%%%%%%%%%%%%%%
%% Closure of the \allowdisplaybreaks command %
%%%%%%%%%%%%%%%%%%%%%%%%%%%%%%%%%%%%%%%%%%%%%%%
\section*{Appendix}
\label{sec: appendix}
\setcounter{equation}{0}
\renewcommand{\theequation}{A.\arabic{equation}}
The formulation of the problem concerning the inflation and extension of an isotropic nonlinear elastic circular cylinder can be found in several textbooks (see, e.g., \citealp{Ogden-NEDBook-1984-1,Humphrey_2010_Cardiovascular_0}).  Aspects of the solution to this problem are also available. However, the authors could not find in the literature the specific expressions they needed in this paper for the Lagrange multiplier, the Cauchy, and second Piola-Kirchhoff stress tensor and their averages through the thickness.  Therefore, we present our derivation of the quantities in question.  We do not claim to have found a new solution to the inflation and extension problem. We have simply derived the expressions for the quantities needed by our analysis.  What is new is our paper is the definition of an effective shear modulus of a tubular vessel that can be measured using current experimental approaches both \emph{in vivo} and \emph{in vitro} and, most importantly, that properly isolates the elastic response of tubular biological vessels from the response to incompressibility.

\subsection*{Inflation and extension of a neo-Hookean tubular vessel}
%Equation~\eqref{Eq: NHs div T = 0} is the only nontrivial component of the balance of linear momentum for the tubular structure in Fig.~\ref{fig: coordinate systems} when modeled as an incompressible isotropic materials with undistorted reference configuration. 
Recalling that $\tensor{I} = \uv{r} \otimes \uv{r} + \uv{\theta} \otimes \uv{\theta} + \uv{z} \otimes \uv{z}$ and that $\tensor{B}$ is given in Eq.~\eqref{Eq: B}, for a new-Hookean material described via Eq.~\eqref{Eq: Effective stress strain}, we have
\begin{equation}
\label{Eq: NH Cauchy stress components rough}
\sigma_{rr} = -q + \frac{\mu}{\lambda_{z}^{2}} \frac{R^{2}}{r^{2}}
\quad \text{and} \quad
\sigma_{\theta\theta} = - q + \mu \frac{r^{2}}{R^{2}}.
\end{equation}
Substituting Eq.~\eqref{Eq: NH Cauchy stress components rough} into Eq.~\eqref{Eq: NHs div T = 0} (the only nontrivial component of the balance of linear momentum for the problem at hand), we have
\begin{equation}
\label{Eq: NH q equilibrium eq}
-\frac{\d q}{\d r} + \frac{\mu}{\lambda_{z}^{2}} \frac{\d}{\d r} \frac{R^{2}}{r^{2}}
+ \frac{\mu}{\lambda_{z}^{2}} \frac{R^{2}}{r^{3}} - \mu \frac{r}{R^{2}} = 0.
\end{equation}
The deformation under consideration allows us to express $r$ as a function of $R$ and \emph{vice versa}.  When we view $R = R(r)$, from the first Eqs.~\eqref{Eq: continuity}, we conclude that $\d R/\d r = \lambda_{z} r / R$.  This allows us to write the last two terms of the left-hand side of Eq.~\eqref{Eq: NH q equilibrium eq} as follows:
\begin{equation}
\label{Eq: continuity related results}
\frac{R^{2}}{r^{3}}
= -\frac{1}{2} \frac{\d}{\d r} \frac{R^{2}}{r^{2}} + \frac{R}{r^{2}} \frac{\d R}{\d r}
= -\frac{1}{2} \frac{\d}{\d r} \frac{R^{2}}{r^{2}} + \lambda_{z} \frac{\d \ln r}{\d r}
\quad \text{and} \quad
\frac{r}{R^{2}} = \frac{1}{\lambda_{z}} \frac{\d \ln R}{\d r}.
\end{equation}
Hence, Eq.~\eqref{Eq: NH q equilibrium eq} can be given the form
\begin{equation}
\label{Eq: NH q equilibrium eq diff}
\frac{\d q}{\d r} = \frac{\mu}{2 \lambda_{z}^{2}} \frac{\d}{\d r} \frac{R^{2}}{r^{2}}
+ \frac{\mu}{\lambda_{z}} \frac{\d}{\d r} \ln \frac{r}{R}
\quad \Rightarrow \quad
q = K + \frac{\mu}{2 \lambda_{z}^{2}} \frac{R^{2}}{r^{2}} + \frac{\mu}{\lambda_{z}} \ln\frac{r}{R},
\end{equation}
where $K$ is a constant of integration that we find by enforcing a boundary condition. The outer boundary of the cylinder, with outward unit normal $\uv{r}$, is subject to the pressure $p_{o}$.  Hence, $\bv{\sigma} \uv{r}|_{r = r_{o}} = -p_{o} \, \uv{r}$, which implies that $\sigma_{rr}(r_{o}) = -p_{o}$.  This expression, along with the first of Eqs.~\eqref{Eq: NH Cauchy stress components rough} and the last of Eqs.~\eqref{Eq: NH q equilibrium eq diff}, yields an equation for $K$ whose solution is
\begin{equation}
\label{Eq: K solution}
K = p_{o} + \frac{\mu R_{o}^{2}}{2\lambda_{z}^{2} r_{o}^{2}} + \frac{\mu}{\lambda_{z}} \ln\frac{R_{o}}{r_{o}}.
\end{equation} 
Substituting Eq.~\eqref{Eq: K solution} into the last of Eqs.~\eqref{Eq: NH q equilibrium eq diff} and simplifying, we have
\begin{equation}
\label{Eq: q final form}
q = 
p_{o} + \frac{\mu}{2\lambda_{z}^{2}} \biggl(\frac{R_{o}^{2}}{r_{o}^{2}} + \frac{R^{2}}{r^{2}} \biggr) + \frac{\mu}{\lambda_{z}} \ln\biggl(\frac{r R_{o}}{r_{o} R}\biggr).
\end{equation}
We now observe that the the inner boundary of the cylinder is subject to the pressure $p_{i}$, which implies that $\sigma_{rr}(r_{i}) = -p_{i}$.  We can use this expression along with Eq.~\eqref{Eq: q final form} to determine an expression relating the transmural pressure difference $\Delta p$ to the elastic constant $\mu$.  Doing so gives
\begin{equation}
\label{Eq: mu delta p rel}
-p_{o} - \frac{\mu}{2\lambda_{z}^{2}} \biggl(\frac{R_{o}^{2}}{r_{o}^{2}} + \frac{R_{i}^{2}}{r_{i}^{2}} \biggr) - \frac{\mu}{\lambda_{z}} \ln\biggl(\frac{r_{i} R_{o}}{r_{o} R_{i}}\biggr)
+ \frac{\mu}{\lambda_{z}^{2}} \frac{R_{i}^{2}}{r_{i}^{2}} = -p_{i}.
\end{equation}
Recalling that $\Delta p = p_{i} - p_{o}$, the above equation can be solved for $\mu$ to obtain
\begin{equation}
\label{Eq: mu solution}
\mu = \Delta p \biggl[\frac{1}{2 \lambda_{z}^{2}} \biggl( \frac{R_{o}^{2}}{r_{o}^{2}} - \frac{R_{i}^{2}}{r_{i}^{2}} \biggr) + \frac{1}{\lambda_{z}} \ln\biggl( \frac{r_{i} R_{o}}{r_{o} R_{i}} \biggr)\biggr]^{-1}.
\end{equation}
The full Cauchy stress solution is then obtained by substituting the expression for $q$ into Eq.~\eqref{Eq: Effective stress strain} and using Eq.~\eqref{Eq: B}.  This gives
\begin{equation}
\label{Eq: sigma final app}
\bv{\sigma} = -
\biggl[p_{o} + \frac{\mu}{2\lambda_{z}^{2}} \biggl(\frac{R_{o}^{2}}{r_{o}^{2}} + \frac{R^{2}}{r^{2}} \biggr) + \frac{\mu}{\lambda_{z}} \ln\biggl(\frac{r R_{o}}{r_{o} R}\biggr)\biggr] \tensor{I}
+ \mu\biggl(
\frac{1}{\lambda_{z}^{2}} \frac{R^{2}}{r^{2}}\, \uv{r} \otimes \uv{r} + \frac{r^{2}}{R^{2}} \, \uv{\theta} \otimes \uv{\theta} + \lambda_{z}^{2} \, \uv{z} \otimes \uv{z}\biggr).
\end{equation}
Using the first of Eqs.~\eqref{Eq: T-S relation and E def}, recalling that in our problem $\det\tensor{F} = 1$, and using Eqs.~\eqref{Eq: F} and~\eqref{Eq: FInv}, we then have that the full solution for the second Piola-Kirchhoff stress tensor is
\begin{equation}
\label{Eq: S final app}
\tensor{S} = -
\biggl[p_{o} + \frac{\mu}{2\lambda_{z}^{2}} \biggl(\frac{R_{o}^{2}}{r_{o}^{2}} + \frac{R^{2}}{r^{2}} \biggr) + \frac{\mu}{\lambda_{z}} \ln\biggl(\frac{r R_{o}}{r_{o} R}\biggr)\biggr] 
\biggl(\lambda_{z}^{2} \frac{r^{2}}{R^{2}} \, \uv{R} \otimes \uv{R}  + \frac{R^{2}}{r^{2}} \, \uv{\Theta} \otimes \uv{\Theta}   + \frac{1}{\lambda_{z}^{2}} \uv{Z} \otimes  \, \uv{Z}\biggr)
%\\
+ \mu\tensor{I}.
\end{equation}

\subsection*{Inflation and extension of a Mooney-Rivlin tubular vessel}
Recalling that $\tensor{B}^{-1}$ is given in Eq.~\eqref{Eq: B inverse}, for a Mooney-Rivlin material described via Eq.~\eqref{Eq: effective MR Material} with $\mu_{1} = 0$, we have
\begin{equation}
\label{Eq: RMR Cauchy stress components rough}
\sigma_{rr} = -q - \mu_{2} \lambda_{z}^{2} \frac{r^{2}}{R^{2}}
\quad \text{and} \quad
\sigma_{\theta\theta} = - q - \mu_{2} \frac{R^{2}}{r^{2}}.
\end{equation}
Substituting Eq.~\eqref{Eq: RMR Cauchy stress components rough} into Eq.~\eqref{Eq: NHs div T = 0} (the only nontrivial component of the balance of linear momentum for the problem at hand), we have
\begin{equation}
\label{Eq: RMR q equilibrium eq}
-\frac{\d q}{\d r} - \mu_{2} \lambda_{z}^{2} \frac{\d}{\d r} \frac{r^{2}}{R^{2}}
- \mu_{2}\lambda_{z}^{2} \frac{r}{R^{2}} + \mu_{2} \frac{R^{2}}{r^{3}} = 0.
\end{equation}
Using Eqs.~\eqref{Eq: continuity related results}, we can rewrite Eq.~\eqref{Eq: RMR q equilibrium eq} as follows:
\begin{equation}
\label{Eq: RMR q diff eq}
\begin{multlined}
%\frac{\d q}{\d r}
%= - \mu_{2} \lambda_{z}^{2} \frac{\d}{\d r} \frac{r^{2}}{R^{2}}
%-\frac{\mu_{2}}{2} \frac{\d}{\d r} \frac{R^{2}}{r^{2}} + \mu_{2} \lambda_{z} \frac{\d }{\d r} \ln \frac{r}{R}
\frac{\d q}{\d r}
= \mu_{2} \lambda_{z}^{2}
\biggl[
\frac{1}{2 \lambda_{z}^{2}} \frac{\d}{\d r} \frac{R^{2}}{r^{2}}
+ \frac{1}{\lambda_{z}} \frac{\d}{\d r} \ln \frac{r}{R}
- \frac{\d}{\d r} \biggl(\frac{r^{2}}{R^{2}} + \frac{1}{\lambda_{z}^{2}} \frac{R^{2}}{r^{2}} \biggr)
\biggr]
\\
\Rightarrow \quad
q
= C + \mu_{2} \lambda_{z}^{2}
\biggl[
\frac{1}{2 \lambda_{z}^{2}} \frac{R^{2}}{r^{2}}
+ \frac{1}{\lambda_{z}} \ln \frac{r}{R}
- \biggl(\frac{r^{2}}{R^{2}} + \frac{1}{\lambda_{z}^{2}} \frac{R^{2}}{r^{2}} \biggr)
\biggr],
\end{multlined}
\end{equation}
where $C$ is a constant of integration that we find by enforcing a boundary condition. Recalling that on the outer boundary of the cylinder we must have$\sigma_{rr}(r_{o}) = -p_{o}$, from the first of Eqs.~\eqref{Eq: RMR Cauchy stress components rough} and the last of Eqs.~\eqref{Eq: RMR q diff eq}, we obtain an equation for $C$ whose solution is
\begin{equation}
\label{Eq: C solution}
%p_{o} - \mu_{2} \lambda_{z}^{2} \frac{r_{o}^{2}}{R_{o}^{2}}
%= C + \mu_{2} \lambda_{z}^{2}
%\biggl[
%\frac{1}{2 \lambda_{z}^{2}} \frac{R_{o}^{2}}{r_{o}^{2}}
%+ \frac{1}{\lambda_{z}} \ln \frac{r_{o}}{R_{o}}
%- \biggl(\frac{r_{o}^{2}}{R_{o}^{2}} + \frac{1}{\lambda_{z}^{2}} \frac{R_{o}^{2}}{r_{o}^{2}} \biggr)
%\biggr],
C = 
p_{o}
+ \mu_{2} \lambda_{z}^{2}
\biggl(
\frac{1}{2 \lambda_{z}^{2}} \frac{R_{o}^{2}}{r_{o}^{2}}
+ \frac{1}{\lambda_{z}} \ln \frac{R_{o}}{r_{o}}
\biggr).
\end{equation}
Substituting Eq.~\eqref{Eq: C solution} into Eq.~\eqref{Eq: RMR q diff eq} and simplifying, we have
\begin{equation}
\label{Eq: q RMR final form}
q = 
p_{o} + \mu_{2} \lambda_{z}^{2}
\biggl[
\frac{1}{2 \lambda_{z}^{2}} \biggl(\frac{R_{o}^{2}}{r_{o}^{2}} + \frac{R^{2}}{r^{2}}\biggr)
+ \frac{1}{\lambda_{z}} \ln \biggl(\frac{r R_{o}}{r_{o}R}\biggr)
- \biggl(\frac{r^{2}}{R^{2}} + \frac{1}{\lambda_{z}^{2}} \frac{R^{2}}{r^{2}} \biggr)
\biggr].
\end{equation}
We now observe that the the inner boundary of the cylinder is subject to the pressure $p_{i}$, which implies that $\sigma_{rr}(r_{i}) = -p_{i}$.  We can use this expression along with Eq.~\eqref{Eq: q  RMR final form} to determine an expression relating the transmural pressure difference $\Delta p$ to the elastic constant $\mu_{2}$.  Doing so gives
\begin{equation}
\label{Eq: mu2 delta p rel}
-p_{o} - \mu_{2} \lambda_{z}^{2}
\biggl[
\frac{1}{2 \lambda_{z}^{2}} \biggl(\frac{R_{o}^{2}}{r_{o}^{2}} + \frac{R_{i}^{2}}{r_{i}^{2}}\biggr)
+ \frac{1}{\lambda_{z}} \ln \biggl(\frac{r_{i} R_{o}}{r_{o}R_{i}}\biggr)
- \biggl(\frac{r_{i}^{2}}{R_{i}^{2}} + \frac{1}{\lambda_{z}^{2}} \frac{R_{i}^{2}}{r_{i}^{2}} \biggr)
\biggr] - \mu_{2} \lambda_{z}^{2} \frac{r_{i}^{2}}{R_{i}^{2}}.
 = -p_{i}.
\end{equation}
Recalling that $\Delta p = p_{i} - p_{o}$, the above equation can be solved for $\mu_{2}$ to obtain
\begin{equation}
\label{Eq: mu2 solution}
\mu_{2} = \frac{\Delta p}{\lambda_{z}^{2}}
\biggl[
\frac{1}{2 \lambda_{z}^{2}} \biggl(\frac{R_{o}^{2}}{r_{o}^{2}} - \frac{R_{i}^{2}}{r_{i}^{2}}\biggr)
+ \frac{1}{\lambda_{z}} \ln \biggl(\frac{r_{i} R_{o}}{r_{o}R_{i}}\biggr)
\biggr]^{-1}.
\end{equation}
The full Cauchy stress solution is then obtained by substituting the expression for $q$ into Eq.~\eqref{Eq: effective MR Material} and using Eq.~\eqref{Eq: B inverse}.  This gives
\begin{equation}
\label{Eq: sigma RMR final app}
\begin{multlined}
\bv{\sigma} = -
\biggl\{
p_{o} + \mu_{2} \lambda_{z}^{2}
\biggl[
\frac{1}{2 \lambda_{z}^{2}} \biggl(\frac{R_{o}^{2}}{r_{o}^{2}} + \frac{R^{2}}{r^{2}}\biggr)
+ \frac{1}{\lambda_{z}} \ln \biggl(\frac{r R_{o}}{r_{o}R}\biggr)
- \biggl(\frac{r^{2}}{R^{2}} + \frac{1}{\lambda_{z}^{2}} \frac{R^{2}}{r^{2}} \biggr)
\biggr]\biggr\} \tensor{I}
\\
- \mu_{2}\biggl(
\lambda_{z}^{2} \frac{r^{2}}{R^{2}}\, \uv{r} \otimes \uv{r} + \frac{R^{2}}{r^{2}} \, \uv{\theta} \otimes \uv{\theta} + \frac{1}{\lambda_{z}^{2}} \, \uv{z} \otimes \uv{z}\biggr).
\end{multlined}
\end{equation}
Using the first of Eqs.~\eqref{Eq: T-S relation and E def}, recalling that in our problem $\det\tensor{F} = 1$, and using Eqs.~\eqref{Eq: F} and~\eqref{Eq: FInv}, we then have that the full solution for the second Piola-Kirchhoff stress tensor is
\begin{equation}
\label{Eq: S RMR final app}
\begin{aligned}
\tensor{S} &= -
\biggl\{
p_{o} + \mu_{2} \lambda_{z}^{2}
\biggl[
\frac{1}{2 \lambda_{z}^{2}} \biggl(\frac{R_{o}^{2}}{r_{o}^{2}} + \frac{R^{2}}{r^{2}}\biggr)
+ \frac{1}{\lambda_{z}} \ln \biggl(\frac{r R_{o}}{r_{o}R}\biggr)
\\
&\qquad\quad- \biggl(\frac{r^{2}}{R^{2}} + \frac{1}{\lambda_{z}^{2}} \frac{R^{2}}{r^{2}} \biggr)
\biggr]\biggr\}
\biggl(\lambda_{z}^{2} \frac{r^{2}}{R^{2}} \, \uv{R} \otimes \uv{R}  + \frac{R^{2}}{r^{2}} \, \uv{\Theta} \otimes \uv{\Theta}   + \frac{1}{\lambda_{z}^{2}} \uv{Z} \otimes  \, \uv{Z}\biggr)
\\
&\qquad\qquad\quad- \mu_{2}\biggl(
\lambda_{z}^{4} \frac{r^{4}}{R^{4}}\, \uv{R} \otimes \uv{R} + \frac{R^{4}}{r^{4}} \, \uv{\Theta} \otimes \uv{\Theta} + \frac{1}{\lambda_{z}^{4}} \, \uv{Z} \otimes \uv{Z}\biggr).
\end{aligned}
\end{equation}

\subsection*{Formulas for the Determination of Averages}
The analysis in the body of the paper uses that average through the vessel's wall of a number of quantities. As it turns out, the determination of these averages consists of combinations of averages of elementary functions.  For the sake of a compact presentation, we limit ourselves to the elementary functions in question and their corresponding averages. 

Let $\rho = \sqrt{r_{i}^{2} - R_{i}^{2}/\lambda_{z}}$.  Recall that $R^{2} = \lambda_{z} r^{2} - \lambda_{z} r_{i}^{2} + R_{i}^{2}$, which can be rewritten as $R^{2} = \lambda_{z}(r^{2} - \rho^{2})$.  The first two elementary functions we consider are
\begin{equation}
\label{Aeq: r2R2 and R2r2}
\frac{R^{2}}{r^{2}} = \lambda_{z} \biggl(1 - \frac{\rho^{2}}{r^{2}}\biggr)
\quad \text{and} \quad
\frac{r^{2}}{R^{2}} = \frac{1}{\lambda_{z}} \biggl( 1 + \frac{\rho^{2}}{r^{2} - \rho^{2}} \biggr).
\end{equation}
Their averages are given by, respectively,
\begin{equation}
\label{Aeq: r2R2 and R2r2 averages}
\biggl\langle \frac{R^{2}}{r^{2}} \biggr\rangle = 
\lambda_{z} \biggl(1 - \frac{\rho^{2}}{r_{o} r_{i}} \biggr)
\quad \text{and} \quad
\biggl\langle \frac{r^{2}}{R^{2}} \biggr\rangle = 
\frac{1}{\lambda_{z}}\biggl[
1 + \frac{\rho}{2(r_{o} - r_{i})}
\ln \frac{(r_{o} - \rho)(r_{i} + \rho)}{(r_{o} + \rho)(r_{i} - \rho)}
\biggr].
\end{equation}
Other three recurring functions are
\begin{equation}
\label{Aeq: R4r4 log1 and rRlog}
\frac{R^{4}}{r^{4}}, \quad
\ln\biggl(
\frac{r R_{o}}{r_{o}R}
\biggr),
\quad \text{and} \quad
\frac{R^{2}}{r^{2}} \ln\frac{R^{2}}{r^{2}}.
\end{equation}
Their averages are, respectively
\begin{align}
\label{Aeq: R4r4 log1 and rRlog avg1}
\bigg\langle
\frac{R^{4}}{r^{4}}
\bigg\rangle
&= \lambda_{z}^{2} - 2 \frac{\lambda_{z}^{2}\rho^{2}}{r_{o}r_{i}} +
\frac{\lambda_{z}^{2} \rho^{4}}{3 r_{o}^{3} r_{i}^{3}} (r_{o}^{2} + r_{o} r_{i} + r_{i}^{2}),
\\
\label{Aeq: R4r4 log1 and rRlog avg2}
\bigg\langle
\ln\biggl(
\frac{r R_{o}}{r_{o}R}
\biggr)
\bigg\rangle
&=
\frac{\rho}{2(r_{o} - r_{i})} \ln \frac{(r_{o} - \rho)(r_{i} + \rho)}{(r_{o} + \rho)(r_{i} - \rho)}
- \frac{r_{i}}{(r_{o} - r_{i})} \ln\frac{r_{i} R_{o}}{r_{o} R_{i}},
\shortintertext{and}
\begin{split}
\label{Aeq: R4r4 log1 and rRlog avg3}
\bigg\langle
\frac{R^{2}}{r^{2}} \ln\frac{R^{2}}{r^{2}}
\bigg\rangle &= \frac{2 \lambda_{z} \rho^{2}}{r_{o} r_{i}}
- \frac{2 \lambda_{z} \rho}{r_{o} - r_{i}} \ln\frac{(r_{o} - \rho)(r_{i} + \rho)}{(r_{o} + \rho)(r_{i} - \rho)}
\\
&\quad+
\frac{\lambda_{z}}{r_{o} - r_{i}}
\biggl[
\biggl(r_{o} + \frac{\rho^{2}}{r_{o}}\biggr) \ln\frac{R_{o}^{2}}{r_{o}^{2}}
-
\biggl(r_{i} + \frac{\rho^{2}}{r_{i}}\biggr) \ln\frac{R_{i}^{2}}{r_{i}^{2}}
\biggr].
\end{split}
\end{align}

\subsection*{Concerning logarithmic averages}
There are a number of expressions that can be written in terms of the principal stretches $\lambda_{r}$, $\lambda_{\theta}$, and $\lambda_{z}$.  In addition, due to the incompressibility constraint, these expressions can be rewritten in terms of $\lambda_{z}$ and either $\lambda_{r}$ or $\lambda_{\theta}$.  When computing logarithmic averages of these expressions, it is often convenient to express integrands in terms of the principal stretch $\lambda_{\theta}$ (and $\lambda_{z}$, although $\lambda_{z}$ appears as a constant in these averages).  In this case, a logarithmic average can be computed as follows:
\begin{equation}
\label{eq: log average change of variables}
\llbracket \bullet \rrbracket =
\frac{1}{\ln(r_{o}/r_{i})} \int_{r_{i}}^{r_{o}} \bullet \, \frac{\d{r}}{r} = \frac{1}{\ln(r_{o}/r_{i})} \int_{\lambda_{\theta_{i}}}^{\lambda_{\theta_{o}}} \bullet \, \frac{1}{\lambda_{\theta}(1 - \lambda_{z} \lambda_{\theta}^{2})} \d{\lambda_{\theta}},
\end{equation}
where
\begin{equation}
\label{eq: log average change of variables limits}
\lambda_{\theta_{i}} = \frac{r_{i}}{R_{i}}
\quad \text{and} \quad
\lambda_{\theta_{o}} = \frac{r_{o}}{R_{o}}.
\end{equation}

}

%%%%%%%%%%%%%%%%%%%%%%%%%%%%%%%%%%%%%%%%%%%%%%%%%%
%: BIBLIOGRAPHY                                  %
%%%%%%%%%%%%%%%%%%%%%%%%%%%%%%%%%%%%%%%%%%%%%%%%%%
\bibliographystyle{FG-AY-bibstyle}                  %
\bibliography{Esophagus} %

\begin{thebibliography}{38}
\newcommand{\enquote}[1]{``#1''}
\providecommand{\natexlab}[1]{#1}
\providecommand{\url}[1]{\texttt{#1}}
\providecommand{\urlprefix}{URL }
\providecommand{\eprint}[2][]{\url{#2}}

\bibitem[Basford(2002)]{Basford_2002_The-Law-of-Laplace_0}
\textsc{Basford, J.~R.} (2002) \enquote{The Law of Laplace and its Relevance to
  Contemporary Medicine and Rehabilitation,} \emph{Archives of Physical
  Medicine and Rehabilitation}, \textbf{83}(8), pp. 1165--1170.

\bibitem[Biancani et~al.(1975)]{BiancaniZabinski_1975_Pressure_0}
\textsc{Biancani, P.}, \textsc{M.~P. Zabinski}, and \textsc{J.~Behar} (1975)
  \enquote{Pressure, Tension, and Force of Closure of the Human Lower
  Esophageal Sphincter and Esophagus,} \emph{Journal of Clinical
  Investigation}, \textbf{56}(2), pp. 476--483.

\bibitem[Bowen(1989)]{Bowen-CMBook-1989-1}
\textsc{Bowen, R.~M.} (1989) \emph{Introduction to Continuum Mechanics for
  Engineers}, vol.~39 of \emph{Mathematical concepts and methods in science and
  engineering}, Plenum Press, New York.

\bibitem[Christensen(1987)]{Christensen_1987_Motor_0}
\textsc{Christensen, J.} (1987) \enquote{Motor Functions of the Esophagus,} in
  \emph{Physiology of the Gastrointestinal Tract} (L.~R. Johnson, ed.), 2nd
  ed., chap.~18, Raven Press, New York.

\bibitem[Dai et~al.(2006)]{DaiKorimilli_2006_Muscle_0}
\textsc{Dai, Q.}, \textsc{A.~Korimilli}, \textsc{V.~K. Thangada}, \textsc{C.~Y.
  Chung}, \textsc{H.~Parkman}, \textsc{J.~G. Brasseur}, and \textsc{L.~S.
  Miller} (2006) \enquote{Muscle Shortening Along the Normal Esophagus During
  Swallowing,} \emph{Digestive Diseases and Sciences}, \textbf{51}(1), pp.
  105--109.

\bibitem[De~Pascalis(2010)]{De-Pascalis_2010_The-Semi-Inverse_0}
\textsc{De~Pascalis, R.} (2010) \emph{The Semi-Inverse Method in Solid
  Mechanics: Theoretical Underpinnings and Novel Applications}, Ph.D. thesis,
  Universit{\'e} Pierre et Marie Curie and Univerisit{\`a} del Salento.

\bibitem[Dodds et~al.(1973)]{DoddsStewart_1973_Movement_0}
\textsc{Dodds, W.~J.}, \textsc{E.~T. Stewart}, \textsc{D.~Hodges}, and
  \textsc{F.~F. Zboralske} (1973) \enquote{Movement of the Feline Esophagus
  Associated with Respiration and Peristalsis,} \emph{The Journal of Clinical
  Investigation}, \textbf{52}(1), pp. 1--13.

\bibitem[Fr{\o}kj{\ae}r
  et~al.(2006)]{FrokjaerAndersen_2006_Ultrasound-determined_0}
\textsc{Fr{\o}kj{\ae}r, J.~B.}, \textsc{S.~D. Andersen}, \textsc{A.~M. Drewes},
  and \textsc{H.~Gregersen} (2006) \enquote{Ultrasound Determined Geometric and
  Biomechanical Properties of the Human Duodenum,} \emph{Digestive Diseases and
  Sciences}, \textbf{51}(9), pp. 1662--1669.

\bibitem[Fung(1993)]{Fung_1981_Biomechanics:_0}
\textsc{Fung, Y.~C.} (1993) \emph{Biomechanics: Mechanical Properties of Living
  Tissues}, 2nd ed., Springer Verlag.

\bibitem[Ghosh et~al.(2008)]{GhoshKahrilas_2008_Liquid_0}
\textsc{Ghosh, S.~K.}, \textsc{P.~J. Kahrilas}, and \textsc{J.~G. Brasseur}
  (2008) \enquote{Liquid in the Gastroesophageal Segment Promotes Reflux, but
  Compliance Does Not: a Mathematical Modeling Study,} \emph{American Journal
  of Physiology - Gastrointestinal and Liver Physiology},
  \textbf{295}({G}920--{G}933).

\bibitem[Ghosh et~al.(2005)]{GhoshKahrilas_2005_The-Mechanical_0}
\textsc{Ghosh, S.~K.}, \textsc{P.~J. Kahrilas}, \textsc{T.~Zaki}, \textsc{J.~E.
  Pandolfino}, \textsc{R.~J. Joehl}, and \textsc{J.~G. Brasseur} (2005)
  \enquote{The Mechanical Basis of Impaired Esophageal Emptying
  Postfundoplication,} \emph{American Journal of Physiology - Gastrointestinal
  and Liver Physiology}, \textbf{289}, pp. {G}21--{G}35.

\bibitem[Gregersen et~al.(2002)]{GregersenGilja_2002_Mechanical_0}
\textsc{Gregersen, H.}, \textsc{O.~H. Gilja}, \textsc{T.~Hausken},
  \textsc{A.~Heimdal}, \textsc{C.~Gao}, \textsc{K.~Matre},
  \textsc{S.~{\O}degaard}, and \textsc{A.~Berstad} (2002) \enquote{Mechanical
  Properties in the Human Gastric Antrum using {B}-Mode Ultrasonography and
  Antral Distension,} \emph{American Journal of Physiology - Gastrointestinal
  and Liver Physiology}, \textbf{283}(2), pp. G368--G375.

\bibitem[Gurtin et~al.(2010)]{GurtinFried_2010_The-Mechanics_0}
\textsc{Gurtin, M.~E.}, \textsc{E.~Fried}, and \textsc{L.~Anand} (2010)
  \emph{The Mechanics and Thermodynamics of Continua}, Cambridge University
  Press, New York.

\bibitem[Heagerty et~al.(1993)]{Heagerty1993Small-Artery-St0}
\textsc{Heagerty, A.~M.}, \textsc{C.~Aalkjaer}, \textsc{S.~J. Bund},
  \textsc{N.~Korsgaard}, and \textsc{M.~J. Mulvany} (1993) \enquote{Small
  Artery Structure in Hypertension. Dual Processes of Remodeling and Growth.}
  \emph{Hypertension}, \textbf{21}(4), pp. 391--397, {doi:
  10.1161/01.HYP.21.4.391}.

\bibitem[Holzapfel(2000)]{Holzapfel-CMBook-2000-1}
\textsc{Holzapfel, G.~A.} (2000) \emph{Nonlinear Solid Mechanics: a Continuum
  Approach for Engineering}, Wiley, Chichester ; New York.

\bibitem[Humphrey(2010)]{Humphrey_2010_Cardiovascular_0}
\textsc{Humphrey, J.~D.} (2010) \emph{Cardiovascular Solid Mechanics---Cells,
  Tissues, and Organs}, Springer-Verlag, New York.

\bibitem[Kassab(2006)]{Kassab2006Biomechanics-of0}
\textsc{Kassab, G.~S.} (2006) \enquote{Biomechanics of the Cardiovascular
  System: the Aorta as an Illustratory Example,} \emph{Journal of The Royal
  Society Interface}, \textbf{3}(11), pp. 719--740, {doi:
  10.1098/rsif.2006.0138}.

\bibitem[Kim et~al.(2002)]{Kim2002Three-Dimension0}
\textsc{Kim, W.~Y.}, \textsc{M.~Stuber}, \textsc{P.~B{\"o}rnert}, \textsc{K.~V.
  Kissinger}, \textsc{W.~J. Manning}, and \textsc{R.~M. Botnar} (2002)
  \enquote{Three-Dimensional Black-Blood Cardiac Magnetic Resonance Coronary
  Vessel Wall Imaging Detects Positive Arterial Remodeling in Patients With
  Nonsignificant Coronary Artery Disease,} \emph{Circulation}, \textbf{106}(3),
  pp. 296--299, {doi: 10.1161/01.CIR.0000025629.85631.1E}.

\bibitem[Miller et~al.(1995)]{MillerLiu_1995_Correlation_0}
\textsc{Miller, L.~S.}, \textsc{J.-B. Liu}, \textsc{F.~P. Colizzo},
  \textsc{H.~Ter}, \textsc{J.~Marzano}, \textsc{C.~Barbarevech},
  \textsc{K.~Hedwig}, \textsc{L.~Leung}, and \textsc{B.~B. Goldberg} (1995)
  \enquote{Correlation of High-Frequency Esophageal Ultrasonography and
  Manometry in the Study of Esophageal Motility,} \emph{Gastroenterology},
  \textbf{109}(3), pp. 832--837.

\bibitem[Mittal et~al.(2006)]{MittalPadda_2006_Synchrony_0}
\textsc{Mittal, R.~K.}, \textsc{B.~Padda}, \textsc{V.~Bhalla},
  \textsc{V.~Bhargava}, and \textsc{J.~Liu} (2006) \enquote{Synchrony between
  Circular and Longitudinal Muscle Contractions during Peristalsis in Normal
  Subjects,} \emph{American Journal of Physiology - Gastrointestinal and Liver
  Physiology}, \textbf{290}, pp. {G}431--{G}438.

\bibitem[Nicosia and Brasseur(2002)]{NicosiaBrasseur_2002_A-Mathematical_0}
\textsc{Nicosia, M.~A.} and \textsc{J.~G. Brasseur} (2002) \enquote{A
  Mathematical Model for Estimating Muscle Tension in vivo during Esophageal
  Bolus Transport,} \emph{Journal of Theoretical Biology}, \textbf{219}(2), pp.
  235--255.

\bibitem[Nicosia et~al.(2001)]{NicosiaBrasseur_2001_Local_0}
\textsc{Nicosia, M.~A.}, \textsc{J.~G. Brasseur}, \textsc{J.-B. Liu}, and
  \textsc{L.~S. Miller} (2001) \enquote{Local Longitudinal Muscle Shortening of
  the Human Esophagus from High-Frequency Ultrasonography,} \emph{American
  Journal of Physiology - Gastrointestinal and Liver Physiology},
  \textbf{281}(4), pp. {G}1022--{G}1033.

\bibitem[Ogden(1997)]{Ogden-NEDBook-1984-1}
\textsc{Ogden, R.~W.} (1997) \emph{Non-Linear Elastic Deformations}, Dover
  Publications, Inc., Mineola, New York.

\bibitem[Pedersen et~al.(2005)]{PedersenDrewes_2005_New-analysis_0}
\textsc{Pedersen, J.}, \textsc{A.~M. Drewes}, and \textsc{H.~Gregersen} (2005)
  \enquote{New Analysis for the Study of the Muscle Function in the Human
  Oesophagus,} \emph{Neurogastroenterology and Motility}, \textbf{17}(5), pp.
  767--772.

\bibitem[Pries et~al.(2005)]{Pries2005Remodeling-of-B0}
\textsc{Pries, A.~R.}, \textsc{B.~Reglin}, and \textsc{T.~W. Secomb} (2005)
  \enquote{Remodeling of Blood Vessels : Responses of Diameter and Wall
  Thickness to Hemodynamic and Metabolic Stimuli,} \emph{Hypertension},
  \textbf{46}(4), pp. 725--731, {doi: 10.1161/01.HYP.0000184428.16429.be}.

\bibitem[Schiffner(2004)]{Schiffner_2004_Opening_0}
\textsc{Schiffner, B.~J.} (2004) \emph{Opening Stiffness of the
  Esophago-Gastric Segment in Health, with {GERD}, and after Endoscopic
  Surgery}, Master's thesis, Department of Mechanical Engineering, The
  Pennsylvania State University.

\bibitem[Shi et~al.(2002)]{ShiPandolfino_2002_Distinct_0}
\textsc{Shi, G.}, \textsc{J.~E. Pandolfino}, \textsc{R.~J. Joehl},
  \textsc{J.~G. Brasseur}, and \textsc{P.~J. Kahrilas} (2002) \enquote{Distinct
  Patterns of Oesophageal Shortening during Primary Peristalsis, Secondary
  Peristalsis and Transient Lower Oesophageal Sphincter Relaxation,}
  \emph{Neurogastroenterology \& Motility}, \textbf{14}(5), pp. 505--512.

\bibitem[Taber(2004)]{Taber_2004_Nonlinear_0}
\textsc{Taber, L.~A.} (2004) \emph{Nonlinear Theory of Elasticity: Application
  in Biomechanics}, World Scientific, River Edge, NJ.

\bibitem[Takeda et~al.(2003)]{TakedaKassab_2003_Effect_0}
\textsc{Takeda, T.}, \textsc{G.~Kassab}, \textsc{J.~M. Liu}, \textsc{T.~Nabae},
  and \textsc{R.~K. Mittal} (2003) \enquote{Effect of Atropine on the
  Biomechanical Properties of the Oesophageal Wall in Humans,} \emph{The
  Journal of Physiology}, \textbf{547}(2), pp. 621--628.

\bibitem[Takeda et~al.(2002)]{TakedaKassab_2002_A-novel_0}
\textsc{Takeda, T.}, \textsc{G.~Kassab}, \textsc{J.~M. Liu}, \textsc{J.~L.
  Puckett}, \textsc{R.~R. Mittal}, and \textsc{R.~K. Mittal} (2002) \enquote{A
  Novel Ultrasound Technique to Study the Biomechanics of the Human Esophagus
  in Vivo,} \emph{American Journal of Physiology - Gastrointestinal and Liver
  Physiology}, \textbf{282}(5), pp. G785--G793.

\bibitem[Takeda et~al.(2004)]{TakedaNabae_2004_Oesophageal_0}
\textsc{Takeda, T.}, \textsc{T.~Nabae}, \textsc{G.~Kassab}, \textsc{J.~Liu},
  and \textsc{R.~K. Mittal} (2004) \enquote{Oesophageal Wall Stretch: the
  Stimulus for Distension Induced Oesophageal Sensation,}
  \emph{Neurogastroenterology and Motility}, \textbf{16}(6), pp. 721--728.

\bibitem[Ulerich et~al.(2003)]{UlerichDai_2003_Detailed_0}
\textsc{Ulerich, R.}, \textsc{Q.~Dai}, \textsc{L.~S. Miller}, and \textsc{J.~G.
  Brasseur} (2003) \enquote{Detailed {3-D} Anatomy of the Human
  Gastro-Esophageal Segment (GES),} \emph{Gastroenterology}, \textbf{124}(4,
  Supplement 1), pp. {A}259--{A}259, abstract S 1747, Digestive Disease Week
  and the 104th Annual Meeting of the American Gastroenterological Association.

\bibitem[{Wolfram Research, Inc.}(2010)]{Wolfram-Research2010Mathematica0}
\textsc{{Wolfram Research, Inc.}} (2010) \emph{Mathematica}, {V}ersion 8.0,
  Wolfram Research Inc{.}, Champaign, Illinois.

\bibitem[Yang et~al.(2004)]{YangZhao_2004_Biomechanical_0}
\textsc{Yang, J.}, \textsc{J.~Zhao}, \textsc{Y.~Zeng}, and
  \textsc{H.~Gregersen} (2004) \enquote{Biomechanical Properties of the Rat
  Oesophagus in Experimental Type-1 Diabetes,} \emph{Neurogastroenterology and
  Motility}, \textbf{16}(2), pp. 195--203.

\bibitem[Yang et~al.(2006{\natexlab{a}})]{YangFung_2006_3D-mechanical_0}
\textsc{Yang, W.}, \textsc{T.~C. Fung}, \textsc{K.~S. Chian}, and \textsc{C.~K.
  Chong} (2006{\natexlab{a}}) \enquote{{3D} Mechanical Properties of the
  Layered Esophagus: Experiment and Constitutive Model,} \emph{Journal of
  Biomechanical Engineering-Transactions of The {ASME}}, \textbf{128}(6), pp.
  899--908.

\bibitem[Yang et~al.(2006{\natexlab{b}})]{YangFung_2006_Directional_0}
---{}---{}--- (2006{\natexlab{b}}) \enquote{Directional, Regional, and Layer
  Variations of Mechanical Properties of Esophageal Tissue and its
  Interpretation Using a Structure-Based Constitutive Model,} \emph{Journal of
  Biomechanical Engineering-Transactions of The {ASME}}, \textbf{128}(3), pp.
  409--418.

\bibitem[Yang et~al.(2006{\natexlab{c}})]{YangFung_2006_Viscoelasticity_0}
---{}---{}--- (2006{\natexlab{c}}) \enquote{Viscoelasticity of Esophageal
  Tissue and Application of a {QLV} Model,} \emph{Journal of Biomechanical
  Engineering-Transactions of The {ASME}}, \textbf{128}(6), pp. 909--916.

\bibitem[Yang et~al.(2007)]{YangFung_2007_Instability_0}
---{}---{}--- (2007) \enquote{Instability of the Two-Layered Thick-Walled
  Esophageal Model under the External Pressure and Circular Outer Boundary
  Condition,} \emph{Journal of Biomechanics}, \textbf{40}(3), pp. 481--490.

\end{thebibliography}
\end{document}